\documentclass[fleqn,usenatbib]{mnras}

\usepackage{newtxtext,newtxmath}

\usepackage[T1]{fontenc}
\usepackage{xspace}

\DeclareRobustCommand{\VAN}[3]{#2}
\let\VANthebibliography\thebibliography
\def\thebibliography{\DeclareRobustCommand{\VAN}[3]{##3}\VANthebibliography}

\usepackage{graphicx}	
\usepackage{amsmath}	
\usepackage{multirow}
\usepackage{subcaption}

\newcommand{\angstrom}{\textup{\AA}\xspace}
\newcommand\textlcsc[1]{\textsc{\MakeLowercase{#1}}}

\newcommand{\lya}{Ly$\alpha$\xspace}
\newcommand{\hb}{H$\beta$\xspace}
\newcommand{\ha}{H$\alpha$\xspace}
\newcommand{\oiilow}{[OII]$\lambda\lambda3726{,}3729$\xspace}
\newcommand{\oiiia}{[OIII]$\lambda4959$\xspace}
\newcommand{\oiiib}{[OIII]$\lambda5007$\xspace}
\newcommand{\oiiiab}{[OIII]$\lambda\lambda4959{,}5007$\xspace}
\newcommand{\niia}{[NII]$\lambda6548$\xspace}
\newcommand{\niib}{[NII]$\lambda6584$\xspace}
\newcommand{\niiab}{[NII]$\lambda\lambda6548{,}6584$\xspace}
\newcommand{\siia}{[SII]$\lambda6716$\xspace}
\newcommand{\siib}{[SII]$\lambda6731$\xspace}
\newcommand{\siiab}{[SII]$\lambda\lambda6716{,}6731$\xspace}

\newcommand{\civ}{CIV$\lambda\lambda1548{,}1551$\xspace}
\newcommand{\heiia}{HeII$\lambda1640$\xspace}
\newcommand{\heiib}{HeII$\lambda4686$\xspace}

\newcommand{\neiv}{[NeIV]$\lambda2422{,}2424$\xspace}

\newcommand{\cii}{[\ion{C}{II}]\xspace}

\newcommand{\hza}{HZ10-E\xspace}
\newcommand{\hzb}{HZ10-C\xspace}
\newcommand{\hzc}{HZ10-W\xspace}

\title[GA-NIFS: Witnessing complex assembly of HZ10]{GA-NIFS: Witnessing the complex assembly of a star-forming system at $\mathbf{z=5.7}$}

\author[G. C. Jones et al.]{
Gareth C. Jones$^{1,2,3}$\thanks{E-mail: gj283@cam.ac.uk},
Andrew J. Bunker$^{1}$,
Kseniia Telikova$^{4}$,
Santiago Arribas$^{5}$,
Stefano Carniani$^{6}$,
\newauthor
Stephane Charlot$^{7}$,
Francesco D’Eugenio$^{2,3}$,
Roberto Maiolino$^{2,3,8}$,
Michele Perna$^{5}$,
Bruno Rodr\'{i}guez Del Pino$^{5}$,
\newauthor
Hannah \"{U}bler$^{2,3,9}$,
Chris Willott$^{10}$,
Manuel Aravena$^{4}$,
Torsten B\"{o}ker$^{11}$,
Giovanni Cresci$^{12}$,
\newauthor
Mirko Curti$^{13}$,
Jorge Gonz\'{a}lez-L\'{o}pez$^{14,15}$,
Rodrigo Herrera-Camus$^{16}$,
Isabella Lamperti$^{5,17,12}$,
\newauthor
Eleonora Parlanti$^{6}$,
Pablo G. P\'erez-Gonz\'alez$^{5}$,
Vicente Villanueva$^{16}$
\\
$^{1}$Department of Physics, University of Oxford, Denys Wilkinson Building, Keble Road, Oxford OX1 3RH, UK\\
$^{2}$Kavli Institute for Cosmology, University of Cambridge, Madingley Road, Cambridge CB3 0HA, UK\\
$^{3}$Cavendish Laboratory, University of Cambridge, 19 JJ Thomson Avenue, Cambridge CB3 0HE, UK\\
$^{4}$Instituto de Estudios Astrof\'{i}sicos, Facultad de Ingenier\'{i}a y Ciencias, Universidad Diego Portales, Av. Ej\'{e}rcito Libertador 441, 8370191 Santiago, Chile\\
$^{5}$Centro de Astrobiolog\'{i}a (CAB), CSIC–INTA, Cra. de Ajalvir Km.~4, 28850- Torrej\'{o}n de Ardoz, Madrid, Spain\\
$^{6}$Scuola Normale Superiore, Piazza dei Cavalieri 7, I-56126 Pisa, Italy\\
$^{7}$Sorbonne Universit\'{e}, CNRS, UMR 7095, Institut d\'Astrophysique de Paris, 98 bis bd Arago, 75014 Paris, France\\
$^{8}$Department of Physics and Astronomy, University College London, Gower Street, London WC1E 6BT, UK\\
$^{9}$Max-Planck-Institut f\"ur extraterrestrische Physik, Gie{\ss}enbachstra{\ss}e 1, 85748 Garching, Germany\\
$^{10}$NRC Herzberg, 5071 West Saanich Rd, Victoria, BC V9E 2E7, Canada\\
$^{11}$European Space Agency, c/o STScI, 3700 San Martin Drive, Baltimore, MD 21218, USA\\
$^{12}$INAF - Osservatorio Astrofisico di Arcetri, largo E. Fermi 5, 50127 Firenze, Italy\\
$^{13}$European Southern Observatory, Karl-Schwarzschild-Strasse 2, 85748 Garching, Germany\\
$^{14}$Instituto de Astrof\'{i}sica, Facultad de F\'{i}sica, Pontificia Universidad Cat\'{o}lica de Chile, Santiago 7820436, Chile\\
$^{15}$Las Campanas Observatory, Carnegie Institution of Washington, Ra\'{u}l Bitr\'{a}n 1200, La Serena, Chile\\
$^{16}$Departamento de Astronom{\'i}a, Universidad de Concepci{\'o}n, Barrio Universitario, Concepci{\'o}n, Chile\\
$^{17}$Dipartimento di Fisica e Astronomia, Universit\`{a} di Firenze\`{a} di Firenze, Via G. Sansone 1, 50019, Sesto Fiorentino, Firenze, Italy
}

\date{Accepted XXX. Received YYY; in original form ZZZ}

\pubyear{2024}

\begin{document}
\label{firstpage}
\pagerange{\pageref{firstpage}--\pageref{lastpage}}
\maketitle

\begin{abstract}
We present observations of the $z\sim5.7$ Lyman-break galaxy HZ10 with the JWST/NIRSpec IFU in high and low spectral resolution (G395H, spectral resolving power $R\sim2700$ and PRISM, $R\sim100$, respectively), as part of the GA-NIFS program. By spatially resolving the source (spatial resolution $\sim0.15''$ or $\sim0.9$\,kpc), we find three spatially and spectrally distinct regions of line emission along with one region of strong continuum emission, all within a projected distance of $<10$\,kpc. The R2700 data features strong detections in \hb, \oiiiab, \niiab, \ha, and \siiab. The R100 data additionally contains a strong detection of the \lya break,  rest-frame UV and optical continuum, and \oiilow. None of the detected lines present strong evidence for AGN excitation from line diagnostic diagrams, and no high-ionisation lines are detected. Using the detected lines, we constrain the electron density $\left( \rm \log_{10}\left( n_e / cm^{-3}\right)\sim 3\right)$ and metallicity ($\sim0.5-0.7$ solar) in each component. Spaxel-by-spaxel fits reveal a strong east-west velocity gradient and significant line asymmetries (possibly indicating tidal features or outflows). The western component features a very red UV slope ($\beta_{\rm UV}\sim-0.9$) and significant \ha emission, suggesting an evolved population and active star formation. A comparison to high resolution  ($\sim0.3''$ or $\sim1.8$\,kpc) \cii\,$158\,\mu$m imaging obtained with the Atacama Large Millimetre/submillimetre Array (ALMA) reveals areas of dust obscuration. Altogether, these data suggest that HZ10 represents an ongoing merger, with a complex distribution of stars, gas, and dust $<1$\,Gyr after the Big Bang.
\end{abstract}




\section{Introduction}

Studies of local galaxies have shown that the majority of them are relaxed disks with spiral arms (e.g., \citealt{delg10}). But when the Universe was much younger ($\lesssim1.5$\,Gyr after the Big Bang, $z>4$), this was not the case. In the early Universe, observational studies have found that the fraction of galaxies with spiral features decreases with increasing redshift (\citealt{kuhn23}), while the major merger fraction increases with redshift (e.g., \citealt{rodr17,dunc19,ferr20}). A survey of $z\sim4-6$ galaxies with the Atacama Large Millimetre/submillimetre Array (ALMA) revealed a small fraction of rotating disks, with a high number of mergers (e.g., \citealt{lefe20,jone21,roma21}; The ALMA Large Program to Investigate \cii at Early Times; ALPINE). This agrees with the results of cosmological zoom-in simulations of high-redshift galaxies, which show that the evolution of these sources is heavily influenced by frequent major and minor mergers (e.g., \citealt{pall17,koha19,kret22}). 

Since the secure identification of merging activity requires resolved observations of galaxy kinematics (e.g., \citealt{simo19}), much progress has been made through ALMA observations of the strong far-infrared (FIR) emission line \cii\,$158\,\mu$m (hereafter \cii). In general, \cii observations of $z>4$ galaxies tend to show multiple star-forming clumps or extensions (e.g., \citealt{cari13,bisc18,carn18,lefe20,nguy20}). Some high spatial resolution ($\sim1-3$\,kpc) \cii observations of these galaxies reveal the presence of ordered rotation (e.g., \citealt{neel20,rizz20,frat21,lell21,tsuk21,herr22,poss23,roma23}), but often with nearby companions or departures from regular rotation that could indicate the influence of a nearby merging companion galaxy. 

These ALMA observations are powerful tools for determining the morpho-kinematics of the cold interstellar medium (ISM) in these galaxies, and may be used to estimate dust properties (from the FIR spectral energy distribution; SED), dynamical mass, ISM properties (e.g., temperature, density, strength of the radiation field; \citealt{nova19,hari20,hash23a}) and either the molecular gas mass (e.g., \citealt{zane18,vizg22a}) or star formation rate (SFR; e.g., \citealt{delo14,scha20}). A complementary view of the ISM and galaxy properties in the context of their morpho-kinematics (including the existence of feedback from the nuclear region; e.g., \citealt{pern20,flue21,cres23}) is enabled through observations of ionised lines in the rest-optical. With the advent of the JWST and the integral field unit (IFU; \citealt{boke22}) on the Near Infrared Spectrograph (NIRSpec; \citealt{jako22}), we may now conduct spatially resolved observations of high-redshift galaxies in rest-UV and rest-optical continuum emission and emission lines, shedding light on stellar populations and the ISM.

The NIRSpec IFU has been used to characterise the morpho-kinematics and ISM conditions for multiple systems of $z>4$ galaxies (e.g., \citealt{hash23b,loia24,vent24}), with a large portion of the observations originating from the Galaxy Assembly with NIRSpec Integral Field Spectroscopy (GA-NIFS\footnote{\url{https://ga-nifs.github.io/}}) 
Guaranteed Time Observations (GTO) program (PIs: R. Maiolino \& S. Arribas; e.g., \citealt{mars23,uble23a,uble24,arri23,uble23b,ji24,jone24,parl23}). This program contains observations of 55 diverse galaxies (e.g., quasar host galaxies, star-forming galaxies, rotators, mergers) between $z\sim2-11$ with the NIRSpec IFU (see also \citealt{deug23a,pern23a,pern23b,pere24,rodr24}). In this work, we analyse data from GA-NIFS observations of the galaxy HZ10 at $z\sim5.7$.

HZ10 was originally detected as a Lyman break galaxy (LBG) in the 2 square degree Cosmic Evolution Survey (COSMOS; \citealt{scov07}) field. Photometric observations with Subaru/Suprime-Cam revealed evidence for significant Ly$\alpha$ emission at $z\sim5.7$ (observed-frame equivalent width of $75\pm22\angstrom$; source 66 of \citealt{mura07}). Follow-up observations with the Deep Extragalactic Imaging Multi-Object Spectrograph (DEIMOS) on the W. M. Keck-II Observatory revealed that this source is a UV-luminous ($M_{\rm UV}=-22.56\pm0.15$) Ly$\alpha$ emitter at $z_{\rm sys}=5.659$, with a red UV slope ($\beta_{\rm UV}<-0.6$) and a large stellar mass ($\log_{10}(M_*/M_{\odot})=10.39\pm0.17$; \citealt{capa15}). It was selected as one of the first high-redshift objects to be observed in \cii emission with ALMA (project 2012.1.00523.S, PI: P. Capak). 

In the sample of ten $z\sim5.1-5.7$ LBGs in \citet{capa15}, HZ10 was the brightest ($S_{\rm 158\mu m}=1.261\pm0.044$\,mJy) and largest (beam-deconvolved FWHM of $(1.16\pm0.11)''\times(0.59\pm0.03)''$ at $83.2\pm3.1^{\circ}$) FIR continuum source. Based on this detection and assuming a modified blackbody model, Capak et al. estimate an infrared luminosity $\log_{10}(L_{\rm IR}/L_{\odot})=11.94\pm0.08$ and $SFR_{\rm IR}=169^{+32}_{-27}$\,M$_{\odot}$\,year$^{-1}$. 
Both the \cii and FIR continuum emission of HZ10 feature an east-west elongation, where the eastern end overlaps with UV emission (Subaru z' band). 

Further analysis of the \cii data by \citet{pave16} returned a precise $z_{\rm \cii}=5.6543\pm0.0003$, $\log_{10}(L_{\rm \cii}/L_{\odot})=9.60\pm0.03$, and the detection of a velocity gradient (FWHM$\sim630$\,km\,s$^{-1}$). While [NII]\,205\,$\mu$m was significantly detected ($>3\sigma$), it featured a smaller full width at half maximum (FWHM), was redshifted, and spatially extended to the west. Based on the observed properties, Pavesi et al. suggest that HZ10 may represent an ongoing major merger. Kinematic modeling of the \cii data cube \citep{pave19} resulted in a dynamical mass estimate of $\log_{10}(M_{\rm dyn}/M_{\odot})=10.79\pm0.05$, in agreement with other analyses (\citealt{capa15,jone17}).

This source has also been detected in CO(2-1) emission \citep{pave19}, where an assumed conversion factor of $\alpha_{\rm CO}=4.5$\,M$_{\odot}$\,K$^{-1}$\,km$^{-1}$\,s\,pc$^{-2}$ yielded a molecular gas mass of $\log_{10}(M_{\rm gas}/M_{\odot})=11.1\pm0.1$. Follow-up observations of CO(5-4) and CO(6-5) \citep{viei22} did not recover significant emission. A detection of CIII]$\lambda1909$ enabled \citet{mark22} to estimate the gas-phase metallicity\footnote{ This metallicity was derived by combining the observed CIII]$\lambda1909$ and \cii fluxes with the model of \citet{vall20}.} of HZ10 ($Z=0.60^{+0.32}_{-0.52}Z_{\odot}$). 

A massive starburst galaxy (CRLE, $z=5.667$) nearby to HZ10 ($13''\sim77$\,kpc in projected distance; $\Delta v\sim580$\,km\,s$^{-1}$) was serendipitously detected through \cii, [NII]205\,$\mu$m, and CO(2-1) observations \citep{pave18}. A further search of the COSMOS2015 photometric redshift catalogue \citep{laig16} revealed the presence of a galaxy overdensity (eight galaxies within $3'$ of CRLE).

Altogether, these previous findings present a complex picture of HZ10. It is gas-rich, dynamically massive, and FIR bright. But its asymmetric velocity field and frequent spatial shift between tracers (e.g., \cii and [NII]) argue against a single source. One possible route towards determining the nature of this galaxy is to examine the rest-frame optical properties of this object on a resolved scale, which was not previously possible. 

In this work we present new JWST/NIRSpec IFU observations of HZ10 to further investigate this well-studied, but still mysterious system. The details of our observations and calibration procedure are given in Section \ref{obscal}. These data are analysed in Section \ref{analysis}, and we discuss the results in Section \ref{discussion}. We conclude in Section \ref{conclusion}. We assume a standard concordance cosmology ($\Omega_{\Lambda}$,$\Omega_m$,h)=(0.7,0.3,0.7) throughout. At the approximate redshift of this source ($z=5.6543$, based on the \cii data of \citealt{capa15} and \citealt{pave16}), $1''$ corresponds to 5.90 projected kpc. Emission lines are named based on their air wavelength, while we use their vacuum wavelengths for analysis (e.g. $\lambda_{\rm rest,[OIII]\lambda5007}=5008.24\angstrom$).

\section{Observations and calibration}\label{obscal}

The data analysed here originated from GA-NIFS observations as part of project 1217 (PI: N. Luetzgendorf; details in Table \ref{JWST_spec}). An eight-point `MEDIUM' dither pattern was used, with a starting point of `1'. Data were taken at both low (PRISM/CLEAR; spectral resolving power $R\sim100$; $0.60-5.30\,\mu$m, hereafter R100) and high spectral resolution (G395H/F290LP; $R\sim2700$; $2.87-5.14\,\mu$m, hereafter R2700). The data were calibrated with the STScI pipeline (v11.1, CRDS 1149), with custom outlier rejection (\citealt{deug23a}), custom masks for cosmic ray strikes (`snowballs') or open MSA shutters, 1/f noise corrections for count-rate maps, and drizzle weighting to create data cubes with spatial pixels (spaxels) of width $0.05''$ (see \citealt{pern23a} for full details of reduction). Previous works targeting bright targets with the JWST/NIRSpec IFU revealed sinusoidal `wiggle' patterns in spectra extracted from single spaxels (e.g., \citealt{pern23b,deca24,uliv24}). These patterns, caused by under-sampling of the PSF, do not appear in our data at high significance (i.e., $>1\sigma$), so we do not apply the correction of \citet{pern23b}.

\begin{table}
\caption{JWST NIRSpec/IFU observation properties.}
\centering
\begin{tabular}{c|cc}
Grating/Filter & G395H/290LP & PRISM/CLEAR \\ \hline
Readout Pattern & NRSIRS2 & NRSIRS2RAPID \\
Groups/Int & 31 & 33 \\
Ints/Exp & 1 & 1 \\
Exposures & 8 & 8 \\
Total Time [ks] & 18.2 & 4.0\\
\end{tabular}
\label{JWST_spec}
\end{table}

To verify the astrometry of our data, we first downloaded HST/WFC3 images from the Mikulski Archive for Space Telescopes (MAST\footnote{\url{https://mast.stsci.edu/portal/Mashup/Clients/Mast/Portal.html}}) archive (F125W, F160W, F105W). These images have been aligned to the \textit{Gaia} DR3 reference frame (\citealt{gaia16,gaia21}), and are well-suited for astrometric comparison. Comparison images were created by convolving the R100 data cube with each HST/WFC3 filter bandpass function. No significant offset (i.e., $>0.1''$) between the HST and JWST images was found, so no astrometric correction was performed. We have also confirmed that the R100 and R2700 cubes are aligned to the same frame (see Appendix \ref{astromapp}). 

No background exposures were included in these observations, so we manually performed a background subtraction for each cube (R100 and R2700) following a procedure similar to that utilised by other NIRSpec IFU studies (e.g., \citealt{scho24,ji25}). First, we mask all data in the cube that falls outside the IFU footprint and apply sigma clipping ($2.5\sigma$) to exclude emission. Using the \textlcsc{photutils} task \textit{Background2D} with a $14\times14$\,spaxel ($0.7''\times0.7''$) square window and a $5\times5$\,spaxel median filter, we determine approximate background maps of each spectral channel in the data cube. These background maps are visually inspected to verify that they do not contain emission (i.e., line or continuum), and are then subtracted from the observed data cube.

\section{Analysis}\label{analysis}

\subsection{Overview of continuum and line emission}\label{oversec}
\begin{figure*}
\centering
\includegraphics[width=\textwidth]{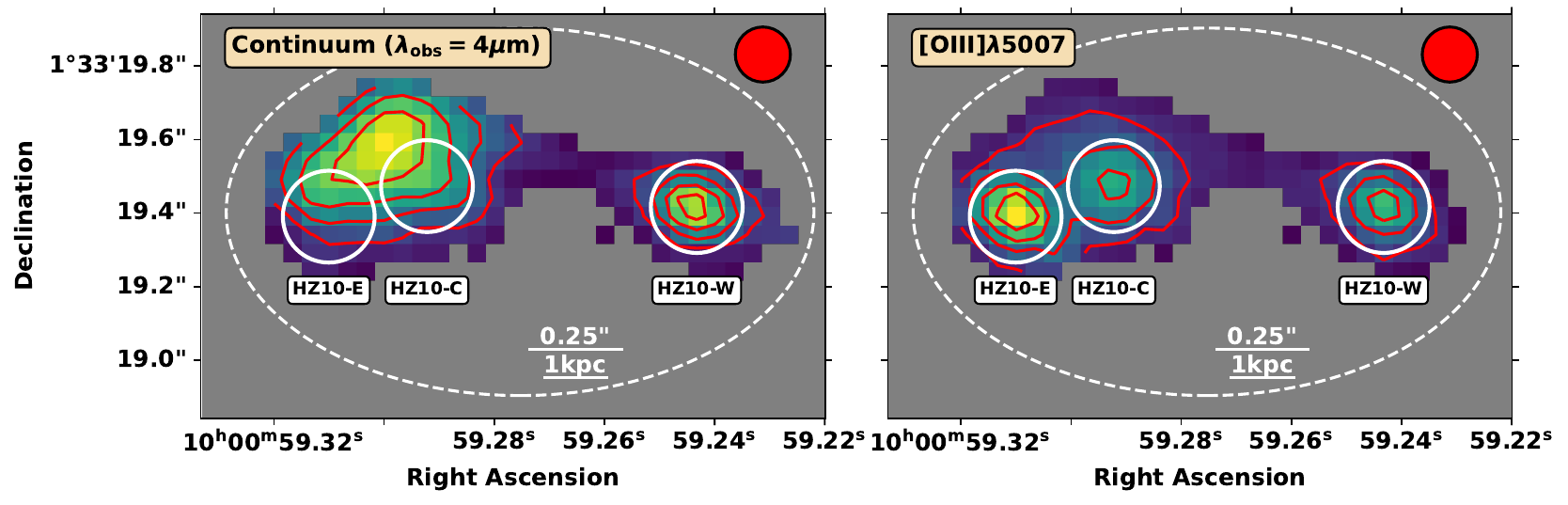} 
\caption{Demonstration of how emission in the HZ10 field is distributed. The colour and red contours show the $\lambda_{\rm obs}=4\,\mu$m continuum level (left panel) and \oiiib emission (right panel) as derived from spaxel-by-spaxel fits to the R2700 data cube (see Section \ref{r2700spx}). Contours are shown at [20, 40, 60, 80]$\%$ of the maximum value of each map. The line emission is concentrated in three regions, which we denote as \hza, \hzb, and \hzc (increasing in redshift and moving from east to west). The apertures we use to extract spectra of each component ($r=0.125''$; Section \ref{CA}) are shown as labelled white circles. A large white dashed ellipse marks the extraction aperture for the analysis of Appendix \ref{ISA}. The fiducial PSF of the data (FWHM=$0.15''$) is shown by a red ellipse to the upper right. Scales of $0.25''$ and 1\,kpc (at a systemic redshift of $z=5.6543$) are shown by white lines. North is up and east is to the left.}
\label{white}
\end{figure*}

An initial investigation of the IFU data cubes revealed that HZ10 is composed of multiple regions of strong emission. To illustrate this, we compare the $\lambda_{\rm obs}=4\,\mu$m continuum (stellar emission at $\lambda_{\rm rest}\sim6000\angstrom$) and \oiiib emission from our R2700 data in Figure \ref{white} (see Section \ref{r2700spx} for details of the creation of this map). It is clear that the line emission is concentrated in three components (here called \hza, \hzb, and \hzc), while the rest-optical continuum is located at the position of \hzc and to the north of \hza and \hzb. We define four apertures for further study: three circular apertures (radius $0.125''$) centred on the line peaks of each component (white circles in Figure \ref{white}) that contain most of the line emission, and one larger ellipse that contains all significant emission (large dashed white ellipse in Figure \ref{white}).

Our ability to detect individual regions is partly due to the narrow point spread function (PSF) of our data (${FWHM}<0.2''$) compared to previous ALMA observations (${FWHM}>0.6''$; \citealt{capa15}). 
\hza and \hzb are only separated by a projected distance of $\sim0.27''$ ($\sim1.6$\,kpc), so previous observations lacked the resolution to separate them (although see novel high-resolution ALMA \cii observations of HZ10; \citealt{teli24,vill24}).

We characterise the emission in this field using two methods. First, we analyse emission from the line peaks (see white circles in Figure \ref{white}; Section \ref{CA}). Next, in Section \ref{RAS} we take advantage of our high spatial resolution by performing spaxel-by-spaxel fits to each data cube. In addition to these analyses, we examine the full field by extracting an integrated spectrum from the R100 data cube using the large dashed white ellipsoidal aperture of Figure \ref{white} (Appendix \ref{ISA})\footnote{Due to the complex morphokinematics of the field (e.g., see Section \ref{r2700spx}), we do not consider the full-field spectrum from the R2700 data cube.}.

\subsection{Component Analysis}\label{CA}

We begin by extracting integrated spectra from the R100 and R2700 cubes using the three emission line-focused apertures and fitting the line and continuum emission. We include \hb, \oiiia, \oiiib, \niia, \ha, \niib, \siia, and \siib in the R2700 fit, while the wider spectral coverage of the R100 data ($\sim0.6-5.2\,\mu$m) allows us to also fit \oiilow. A corrected error spectrum is derived by rescaling the pipeline-output error array to match the standard deviation of sigma-clipped data (e.g., \citealt{uble23b,jone24}). For R100, we derive the scaling factor using data in a spectrally flat region ($\sim4\,\mu$m).

We adopt the standard assumptions of \niib/\niia$=2.94$ (e.g. \citealt{dojc23}) and \oiiib/\oiiia$=2.98$ (e.g. \citealt{dimi07}) for each fit. The ratio \siia/\siib is a strong tracer of electron density, and is restricted to a range of $\sim0.55-1.45$, assuming $T_e=10^4$\,K (e.g., 
\citealt{leti11,dell20}). Due to the low spectral resolution, we model \oiilow as a single Gaussian.

For the R2700 spectra, we allow each line to contain a narrow and broad model component at different redshifts ($z_{\rm N}$ for all narrower components, and $z_{\rm B}$ for all broader components). This represents a two-zone model, where the narrow emission originates from core galaxies and the broad component may originate from nearby tidal features or outflows. This assumption is inspected further in Section \ref{r2700spx} and Appendix \ref{r2700mom}.

To allow for the suppression of \oiiiab and other forbidden lines in high-density environments (see Section \ref{blrsec}), we separate the FWHM values of the Balmer lines (\ha and \hb) from the other lines. That is, we fit four FWHM values (in km\,s$^{-1}$) for each extracted spectrum: the narrow and broad FWHMs for the Balmer lines, and the narrow and broad FWHMs for the other lines. We account for the effect of the spectral resolving power on each linewidth when calculating each FWHM in the R2700 spectrum by adding the modeled intrinsic linewidth and instrumental line spread function (LSF\footnote{As recorded in the JWST documentation; \url{https://jwst-docs.stsci.edu/jwst-near-infrared-spectrograph/nirspec-instrumentation/nirspec-dispersers-and-filters}}) in quadrature. 

The line widths of the R100 fits are set using the fiducial LSF (i.e., each line is assumed to be unresolved in velocity space). Previous investigations using the micro-shutter assembly (MSA) on NIRSpec have found that the true LSF is narrower than the fiducial profile (e.g., \citealt{degr24,jone24}). This is due to the fact that the fiducial spectral resolving power curve was derived assuming a uniformly lit MSA slit (width $0.2''$), while many targets did not fill the slit. However, the IFU constructs data cubes using an image slicer with a narrower width ($0.1''$), so the fiducial LSF should be appropriate for our analysis. Regardless, the LSF-deconvolved width of the R100 lines is not used in this work. Future works will investigate the LSF of the IFU in more detail.

The continuum of the R2700 spectrum is taken into account by including a single power law in the model with slope $\alpha_{\rm opt}^{R2700}$. For the R100 spectrum, the \lya break was modeled using a Heaviside function. To allow for a discontinuity around the Balmer break, we fit the continuum redwards and bluewards of a Balmer series line (H$\eta$, $\lambda_{\rm rest}=3837\angstrom$) with separate power law models (slopes $\alpha_{\rm opt}^{R100}$ and $\beta_{\rm UV}^{R100}$). Previous studies have found that the continuum redwards of the \lya break may deviate from the single power law seen at rest-UV wavelengths (e.g., \citealt{came23b,jone23}). This effect is accounted for by fitting the continuum blue-wards of a pivot wavelength ($\lambda_{\rm rest}=0.15\,\mu$m) with a separate power law component (slope $\beta_{\rm Ly\alpha}^{R100}$), such that the model is continuous at $\lambda_{\rm rest}=0.15\,\mu$m. Due to the low spectral resolution of these data, we convolve the continuum model with a wavelength-dependent Gaussian representing the LSF (e.g., \citealt{jone24,umed23,hein23,napo24}).

\begin{figure*}
\centering
\includegraphics[width=\textwidth]{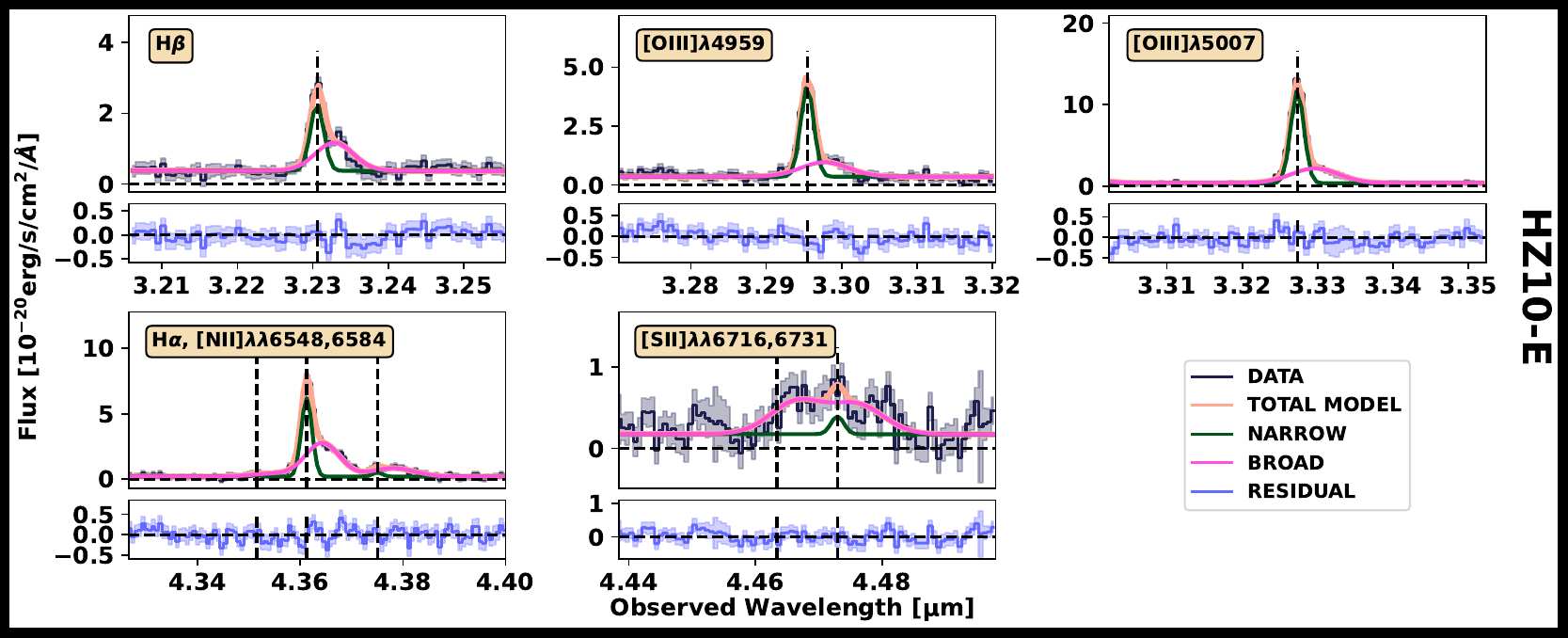}
\includegraphics[width=\textwidth]{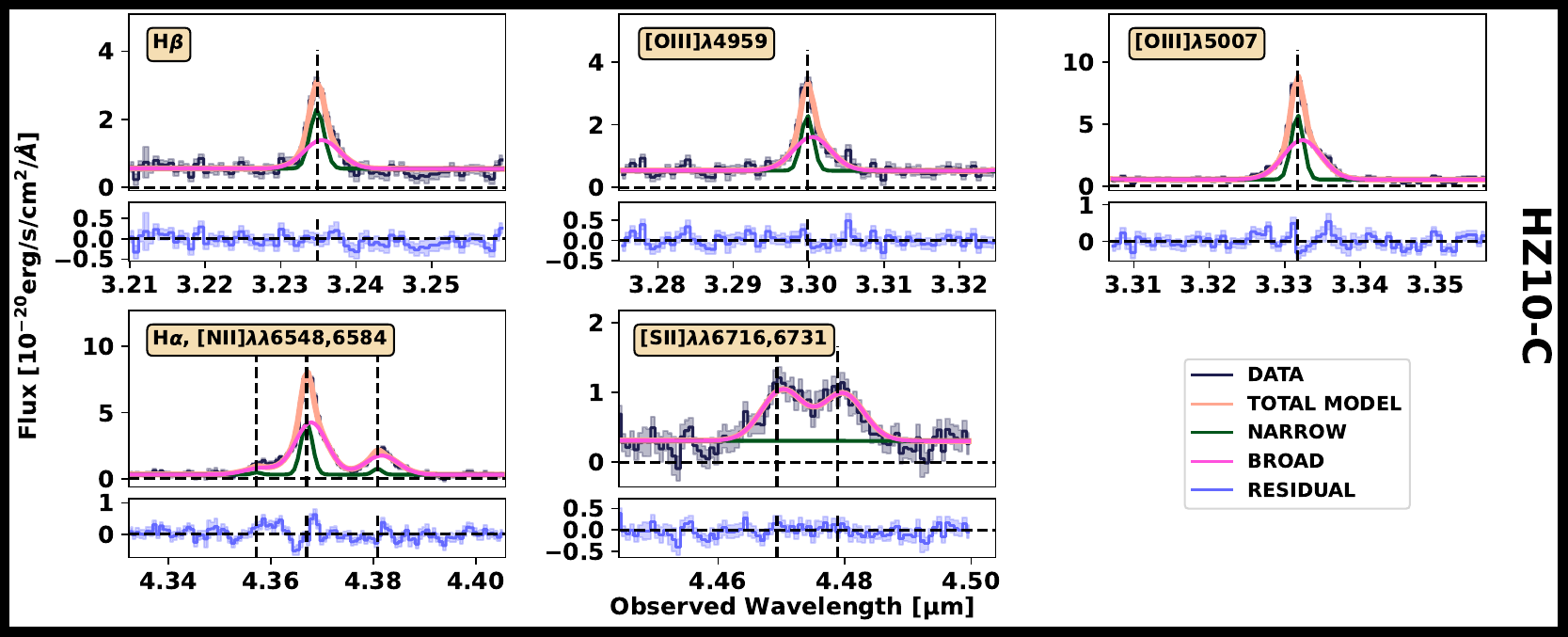}
\includegraphics[width=\textwidth]{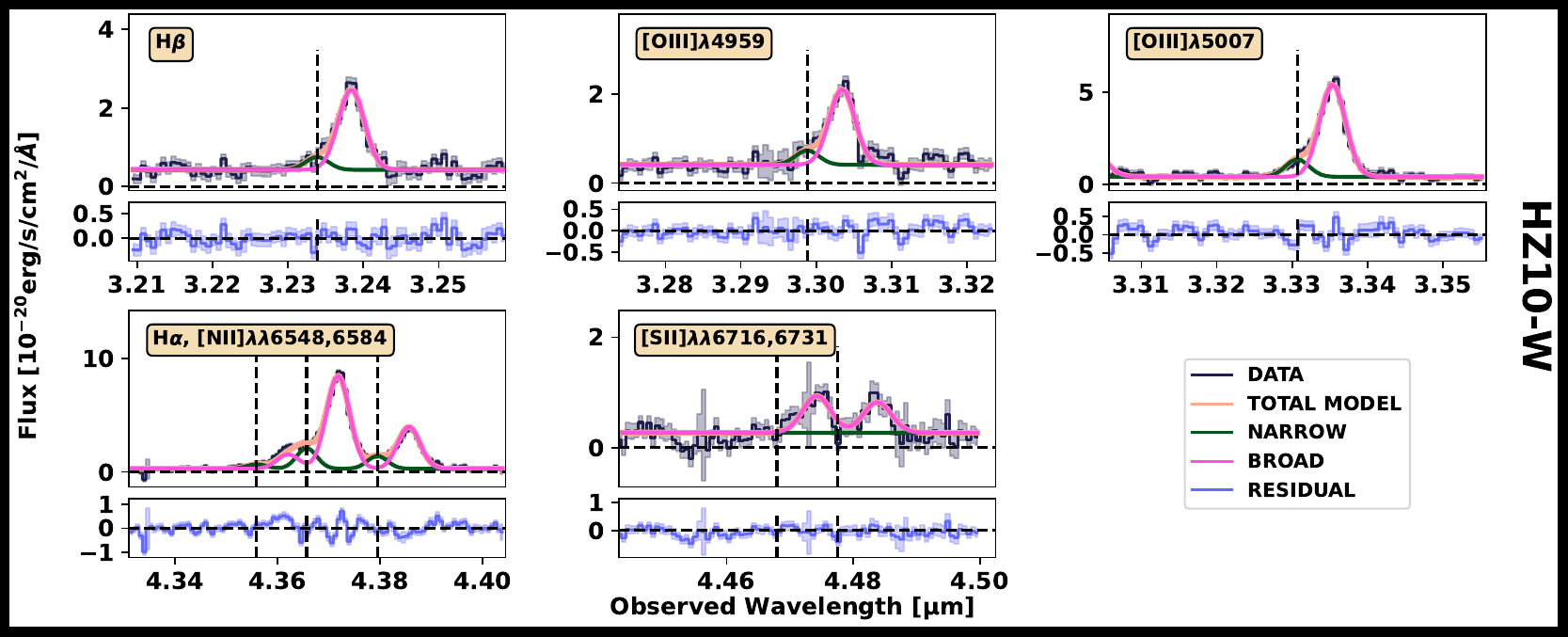}
\caption{Spectra extracted from R2700 data cube using the circular apertures of Figure \ref{white} for each component (black lines). The best-fit model is shown (tan lines), as well as the best-fit narrow (green lines) and broad contributions (pink lines). Model residuals are shown in the lower portion of each panel (blue lines). The best-fit centroid wavelengths of the narrow components of each line are shown by dashed vertical lines. We only show the wavelength range around studied emission lines. Uncertainties ($1\sigma$) for the extracted spectrum and residuals are shown by shaded regions.}
\label{CA_2700}
\end{figure*}

\begin{figure*}
\centering
\includegraphics[width=\textwidth]{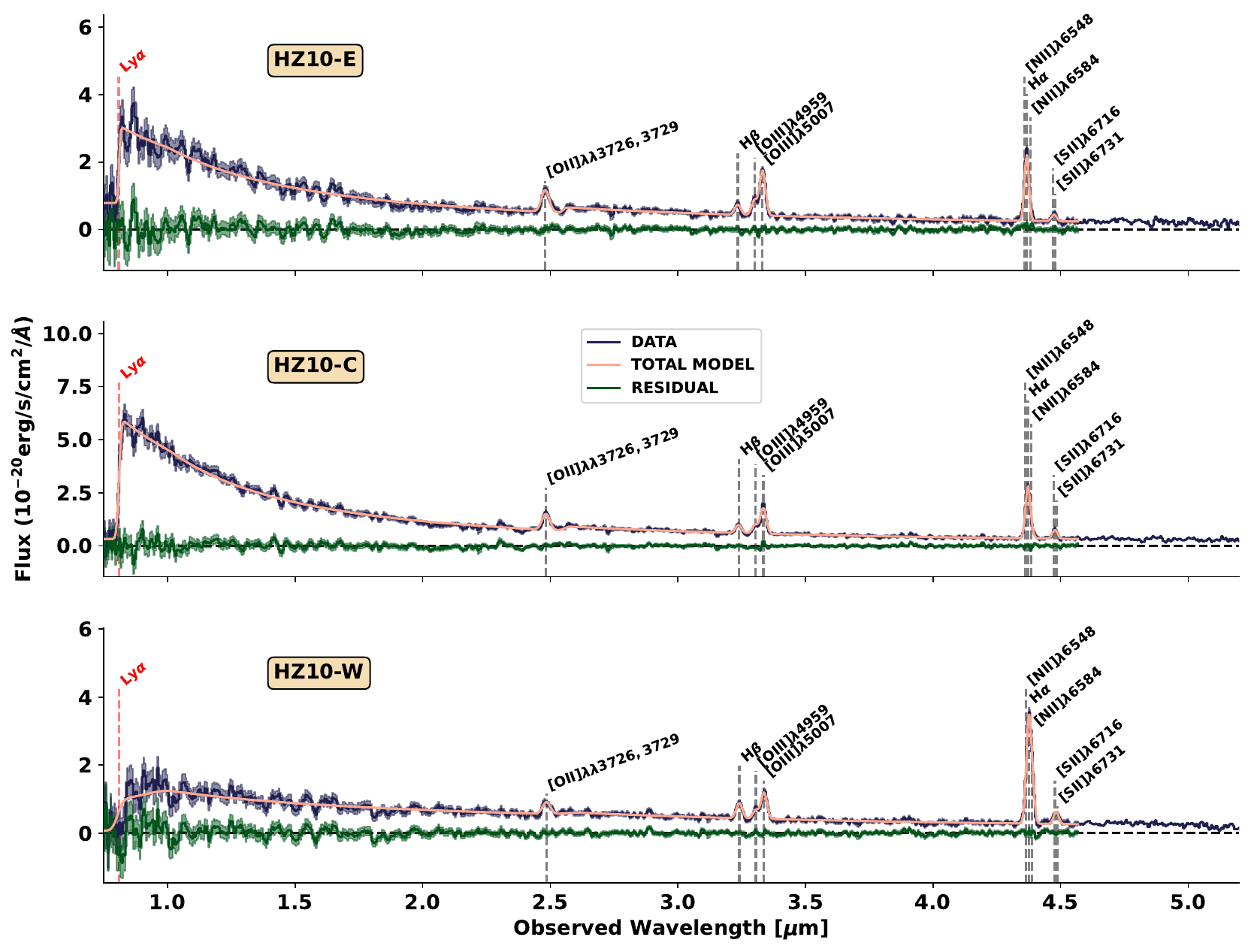}
\caption{Spectra extracted from R100 data cube using the circular apertures of Figure \ref{white} for each component (black lines). The best-fit models are shown by tan lines, as well as the residuals (green lines). The wavelengths of each line (using the best-fit $z_{\rm sys}$) are shown by dashed vertical lines. The expected wavelength of \lya (which is not detected) is marked with a dashed red line. Uncertainties ($1\sigma$) for the extracted spectrum and residuals are shown by shaded regions.}
\label{CA_100}
\end{figure*}

Each spectrum is fit with LMFIT \citep{newv14} in `least\_squares' mode, using inverse variance weighting. The resulting best-fit models are shown in Figure \ref{CA_2700} for R2700 data and Figure \ref{CA_100} for R100 data. The best-fit parameters are presented in Table \ref{linetable_CA}. 

\begin{table*}
\caption{Best-fit line properties for HZ10, derived through separate fits to spectra extracted from $r=0.125''$ circular apertures centred on each component (see Figures \ref{CA_2700} and \ref{CA_100}). Results are listed for the R100 (first set of rows) and R2700 spectra (second set of rows) using multi-Gaussian and continuum models. FWHM values are given in km\,s$^{-1}$, while integrated fluxes are in $10^{-20}$erg\,s$^{-1}$\,cm$^{-2}$. {For definitions of spectral slopes, see Section \ref{CA}.} The R100-based $z_{\rm sys}$ value is not used in this analysis. Velocity offsets are given with respect to $z_{\rm \cii}=5.6543$ \citep{pave16}. We present best-fit fluxes for each line for the R100 and R2700 data. For the R2700 data, we additionally list the best-fit narrow and broad components of the best-fit models. Output uncertainties are given as $1\sigma$, and upper limits are given at $3\sigma$.}
\centering
\begin{tabular}{c|ccc}
	&	HZ10-E	&	HZ10-C	&	HZ10-W	\\	\hline
$z_{\rm sys}$	&	$5.6528\pm0.0006$	&	$5.6593\pm0.0006$	&	$5.6645\pm0.0007$	\\
$M_{\rm UV}$	&	$-19.97\pm0.03$	&	$-20.63\pm0.01$	&	$-19.26\pm0.05$	\\
$\alpha_{\rm opt}^{R100}$	&	$-1.76\pm0.1$	&	$-1.79\pm0.07$	&	$-1.48\pm0.11$	\\
$\beta_{\rm UV}^{R100}$	&	$-1.69\pm0.02$	&	$-1.96\pm0.01$	&	$-0.88\pm0.04$	\\
$\beta_{\rm Ly\alpha}^{R100}$	&	$-1.11\pm0.07$	&	$-1.38\pm0.04$	&	$1.34\pm0.32$	\\
$I_{\rm [OII]\lambda\lambda3726,3729}$	&	$243\pm22$	&	$279\pm21$	&	$130\pm23$	\\	\hline
$z_{\rm N}$	&	$5.6436\pm0.0001$	&	$5.6524\pm0.0001$	&	$5.6504\pm0.0002$	\\
$z_{\rm B}$	&	$5.6483\pm0.0002$	&	$5.6537\pm0.0001$	&	$5.6597\pm0.0001$	\\
$v_{\rm N}$	&	$-480\pm13$	&	$-84\pm13$	&	$-174\pm17$	\\
$v_{\rm B}$	&	$-269\pm17$	&	$-25\pm14$	&	$245\pm13$	\\
$FWHM_{\rm Balmer,N}$	&	$136\pm6$	&	$151\pm13$	&	$300\pm29$	\\
$FWHM_{\rm Balmer,B}$	&	$528\pm17$	&	$492\pm14$	&	$347\pm8$	\\
$FWHM_{\rm [OIII],N}$	&	$129\pm4$	&	$107\pm13$	&	$300\pm40$	\\
$FWHM_{\rm [OIII],B}$	&	$655\pm28$	&	$510\pm19$	&	$334\pm10$	\\
$\alpha_{\rm opt}^{\rm R2700}$	&	$-2.32\pm0.23$	&	$-1.81\pm0.15$	&	$-1.4\pm0.18$	\\
$I_{\rm H\beta,N}$	&	$42\pm3$	&	$41\pm5$	&	$12\pm3$	\\
$I_{\rm H\beta,B}$	&	$50\pm4$	&	$49\pm8$	&	$88\pm4$	\\
$I_{\rm H\beta}$	&	$92\pm5$	&	$90\pm9$	&	$100\pm5$	\\
$I_{\rm [OIII]\lambda4959,N}$	&	$84\pm2$	&	$36\pm3$	&	$12\pm2$	\\
$I_{\rm [OIII]\lambda4959,B}$	&	$48\pm3$	&	$67\pm3$	&	$72\pm2$	\\
$I_{\rm [OIII]\lambda4959}$	&	$132\pm3$	&	$102\pm4$	&	$84\pm2$	\\
$I_{\rm [OIII]\lambda5007,N}$	&	$249\pm4$	&	$106\pm8$	&	$37\pm5$	\\
$I_{\rm [OIII]\lambda5007,B}$	&	$143\pm9$	&	$199\pm10$	&	$215\pm6$	\\
$I_{\rm [OIII]\lambda5007}$	&	$393\pm10$	&	$305\pm13$	&	$252\pm7$	\\
$I_{\rm [NII]\lambda6548,N}$	&	$3\pm1$	&	$3\pm1$	&	$19\pm2$	\\
$I_{\rm [NII]\lambda6548,B}$	&	$23\pm2$	&	$40\pm2$	&	$68\pm2$	\\
$I_{\rm [NII]\lambda6548}$	&	$25\pm2$	&	$43\pm2$	&	$87\pm3$	\\
$I_{\rm H\alpha,N}$	&	$157\pm6$	&	$111\pm11$	&	$89\pm8$	\\
$I_{\rm H\alpha,B}$	&	$214\pm9$	&	$307\pm10$	&	$461\pm9$	\\
$I_{\rm H\alpha}$	&	$371\pm11$	&	$418\pm15$	&	$550\pm12$	\\
$I_{\rm [NII]\lambda6584,N}$	&	$8\pm3$	&	$10\pm4$	&	$56\pm6$	\\
$I_{\rm [NII]\lambda6584,B}$	&	$67\pm6$	&	$116\pm6$	&	$201\pm7$	\\
$I_{\rm [NII]\lambda6584}$	&	$75\pm7$	&	$126\pm7$	&	$257\pm9$	\\
$I_{\rm [SII]\lambda6716,N}$	&	$<37$	&	$<33$	&	$<26$	\\
$I_{\rm [SII]\lambda6716,B}$	&	$42\pm6$	&	$60\pm5$	&	$37\pm5$	\\
$I_{\rm [SII]\lambda6716}$	&	$42\pm6$	&	$60\pm5$	&	$37\pm5$	\\
$I_{\rm [SII]\lambda6731,N}$	&	$6\pm3$	&	$<27$	&	$<27$	\\
$I_{\rm [SII]\lambda6731,B}$	&	$37\pm7$	&	$56\pm5$	&	$31\pm6$	\\
$I_{\rm [SII]\lambda6731}$	&	$43\pm8$	&	$56\pm5$	&	$31\pm6$	\\
\end{tabular}
\label{linetable_CA}
\end{table*}

\subsubsection{Spectral fit results}
The spectra of all three components are well-fit with our models, resulting in low residuals and estimates of redshift, UV magnitude, spectral slopes, line-widths, and line fluxes. 

We are able to measure $z_{\rm sys}$ from the R100 spectra, as well as the redshifts of the narrow ($z_{\rm N}$) and broad contributions ($z_{\rm B}$) in the R2700 spectra. For all components, the R100-based redshift is $>3\sigma$ higher than the average of the narrow and broad R2700-based redshifts ($z_{\rm R100}-z_{\rm R2700}=0.006-0.009$). To explain this, we note that recent NIRSpec studies using the MSA (e.g., \citealt{deug24}) have found a spectral offset between R100 and R1000 data due to NIRSpec wavelength calibration issues ($z_{\rm R100}-z_{\rm R1000}\sim 0.004-0.005$; \citealt{bunk24,deug24,jone25}). Since this affects all NIRSpec observations, IFU data may be affected as well, as confirmed by the NIRSpec/IFU analysis of \citet{pere24}. These authors measured a redshift difference between the R100 and R2700 data for Hyde ($z_{\rm R100}-z_{\rm R2700}=0.0075$), which is larger than the R100-R1000 redshift offset measured for NIRSpec/MSA observations but is comparable to our measured redshift offsets for HZ10. Because the R2700 redshift is in better agreement with the ALMA-based redshift, we only use redshifts based on the R2700 data in this work.

Taking the \cii redshift of \citet{pave16} as the systemic redshift, we see that all three components feature blueshifted narrow emission with broad emission that is relatively redshifted (see Table \ref{linetable_CA}). The line emission from the broad component of \hzc is much stronger than that of the narrow component, possibly indicating outflows or tidal interactions (see Section \ref{hzcsec}). The total velocity difference between the narrow components of \hza and \hzc is large ($\sim660$\,km\,s$^{-1}$), but the difference between the other component pairs is $<500$\,km\,s$^{-1}$.

None of the R100 spectra show evidence for significant \lya emission. Since the previously reported detection of \lya in HZ10 \citep{mura07} used a larger aperture, we also search for \lya using the dashed ellipse in Figure \ref{white} (see analysis of Appendix \ref{ISA}). Even with a broader aperture, we do not detect \lya. However, our $3\sigma$ upper limit on the rest-frame equivalent width of \lya is in agreement with that of \citet{mura07}, suggesting that higher sensitivity observations are required.

The best-fit rest-optical slopes ($\alpha_{\rm opt}$) of the R100 and R2700 data are in agreement ($<3\sigma$ discrepant). The best-fit power law slopes bluewards of $\lambda_{\rm rest}=1500\angstrom$ ($\beta_{\rm Ly\alpha}^{100}$) are all more red than $\beta_{\rm UV}^{100}$, which may be caused by a number of physical conditions (e.g., nebular or two-photon continuum, AGN activity, damped \lya absorbers; \citealt{came23b,tacc24, wits24}). Of the three components, \hzb is the most UV-bright (M$_{\rm UV}=-20.67\pm0.01$) and features the bluest continuum slope ($\beta_{\rm UV}^{100}=-1.99\pm0.01$). On the other hand, \hzc is the most UV-faint (M$_{\rm UV}=-19.4\pm0.04$) and features the reddest continuum slope ($\beta_{\rm UV}^{100}=-1.04\pm0.03$), and \hza is intermediate in both measures. This is discussed further in Section \ref{natco}.

\subsubsection{Line ratio analysis}\label{LRA}
Using the line fluxes from the spectral analysis detailed in the previous subsection, we next explore the position of these components on line ratio diagrams (see Table \ref{lineRtable} for definitions of ratios used in this work). These include the [NII]-BPT ($\log_{10}$R3 vs $\log_{10}$N2; \citealt{bald81}) and [SII]-VO87 ($\log_{10}$R3 vs $\log_{10}$S2; \citealt{veil87}) diagrams. Historically, these plots have been used to determine if galaxies are primarily powered by star formation (low N2, R3, and S2), AGN (high N2, R3, and S2), or a composite of the two (intermediate N2 and R3; e.g., \citealt{ibar15,mazz21,rich21}). While the standard demarcation lines (\citealt{kewl01,kauf03}) are suitable for high-metallicity, massive systems, they were derived for low-redshift galaxies. Some models and observations suggest that these criteria are not suitable for low-metallicity sources at high-redshift ($z>4$; e.g., \citealt{felt16,naka22,maio23b,uble23a,dors24}). Indeed, the location of galaxies on these diagrams is affected by a number of parameters (e.g., $Z$, presence of shocks, ionisation parameter, hydrogen density; \citealt{alle08,felt16,suga21}), so additional diagnostics are required for concrete conclusions. In this work, we utilise the demarcation lines of \citet{schol23}, which were derived from JWST/NIRSpec observations to separate extreme AGN (above each line) from all other sources at high redshift (including both weaker AGN and SFGs; below each line). 

\begin{table}
\centering
\begin{tabular}{c|c}
Ratio & Definition \\ \hline
R3 & \oiiib/\hb \\
N2 & \niib/\ha \\
S2 & \siiab/\ha \\
RS32 & \oiiib/\hb+\siiab/\ha \\
O3S2 & \oiiib/\hb\,/\,\siiab/\ha \\
O3N2 & \oiiib/\hb\,/\,\niib/\ha
\end{tabular}
\caption{Definitions of line flux ratios used in this work.}
\label{lineRtable}
\end{table}

\begin{figure}
\centering
\includegraphics[width=0.5\textwidth]{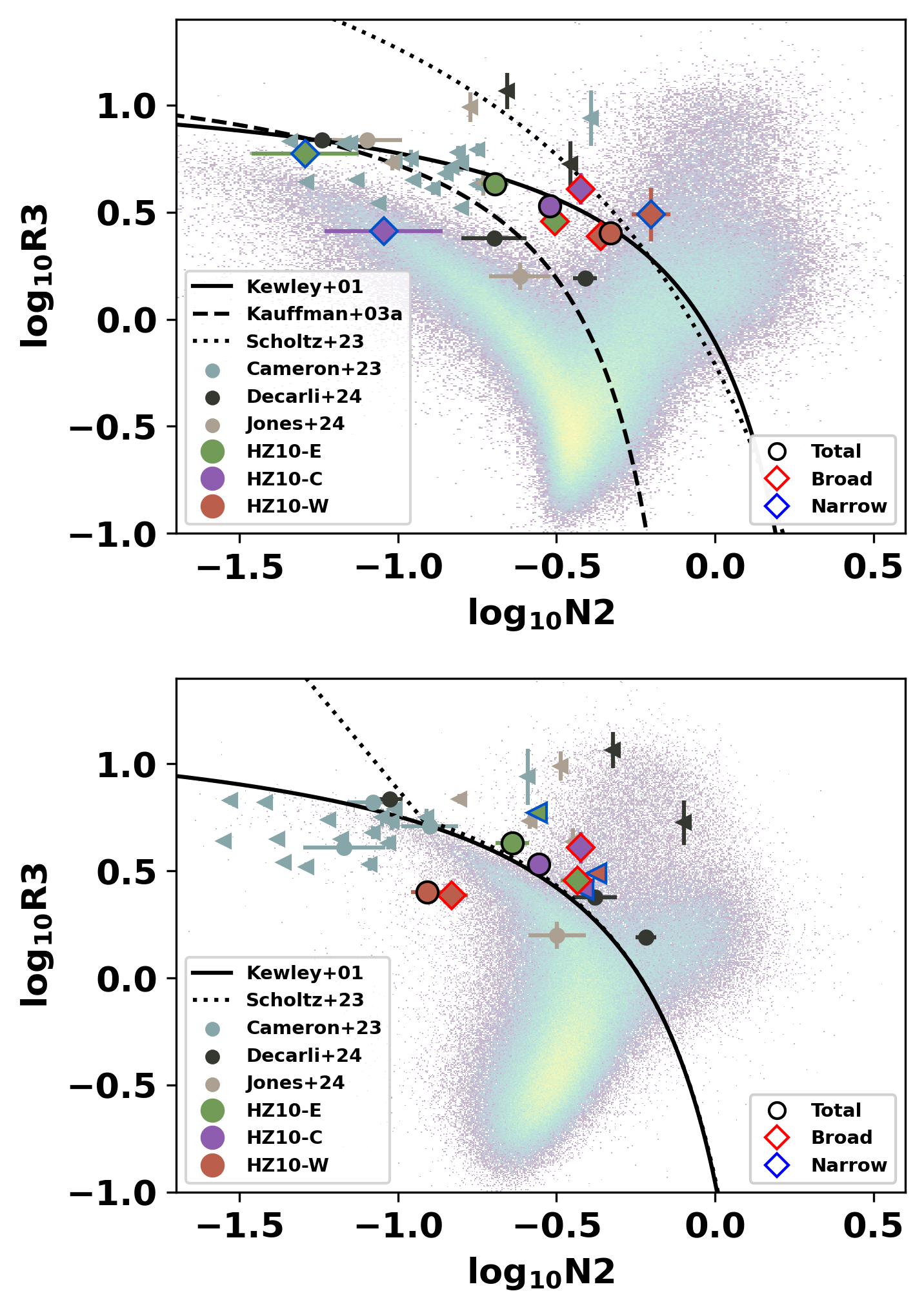}
\caption{[NII]-BPT (upper) and [SII]-VO87 (lower) plots created using best-fit line fluxes for each component of HZ10, as derived from the R2700 spectra (see the values in Table \ref{linetable_CA}). Best-fit narrow components, broad components, and total models for each component are shown with blue, red, and black outlines, respectively. The distributions of low-redshift galaxies from SDSS (MPA-JHU DR8 catalogue; \citealt{kauf03sdss,brin04}) are shown as background points. Left-facing arrows represent $3\sigma$ upper limits. We include the demarcation lines for the $z\sim0$ galaxies of \citet[solid lines]{kewl01} and \citet[dashed line]{kauf03}, as well as the high-redshift demarcation of \citet[dotted line]{schol23}. We compare our results with values from $z\sim5.5-7.0$ galaxies observed with the JWST/NIRSpec MSA (\citealt{came23a}) as well as JWST/NIRSpec IFU observations of the $z=6.2342$ quasar system  PJ308–21 (\citealt{deca24}) and the $z=6.34-6.36$ galaxies in HFLS3 (also known as 1HERMES S350
J170647.8+584623; \citealt{riec13,riec22,jone24}).}
\label{BPT}
\end{figure}

The resulting diagnostic plots are presented in Figure \ref{BPT}. Due to the construction of our models, we may examine the flux ratios of the best-fit narrow (blue outlined symbols), broad (red outlined symbols), and total model (black outlined symbols) for each component. Using the best-fit line fluxes from fits of the R2700 data, we find that the components of HZ10 feature similar line ratios to other observed $z=5.5-7.0$ galaxies (\citealt{came23a,deca24,jone24}). All components fall underneath the \citet{schol23} line for the [NII]-BPT diagram (except for the narrow component of \hzc), but both of the total best-fit models for \hza and \hzc lie on the demarcation line for the [SII]-VO87 relation. The broad components and upper limits on the narrow components of \hzb and \hza lie above the [SII]-VO87 relation demarcation line, adding slight evidence for the presence of an AGN. Because none of the components lie far above the demarcation lines in both plots, there is no strong evidence for the presence of an AGN, although AGN contribution cannot be ruled out (see Section \ref{agnbigsec}). 

The observed line ratios may be used to determine the dust-corrected $SFR_{\rm H\alpha}$. We first calculate the B-V colour excess from the Balmer decrement (e.g. \citealt{domi13,jone24}). Combined with an assumed intrinsic value of $F_{\rm H\alpha}/F_{\rm H\beta}=2.86$ \citep{oste89}\footnote{This ratio was derived assuming Case B recombination, which has recently been called into question for some galaxies (e.g., \citealt{mccl24,scar24}). Since our observed Balmer ratios may be explained with dust attenuation (see Table \ref{ratvals}), this assumption appears to be valid here.}, we correct the observed $F_{\rm H\alpha}$ for dust obscuration and use the \citet{kenn98} scaling law to predict the dust-corrected $SFR_{\rm H\alpha}$ (see Table \ref{ratvals}). The sum of dust-corrected $SFR_{\rm H\alpha}$ values over all three components ($150\pm18$\,M$_{\odot}$\,year$^{-1}$) contains contributions from both obscured and unobscured star formation activity, and is in agreement with the obscured SFR of \citet{capa15}: $SFR_{\mathrm{IR}}=169_{-27}^{+32}\,$M$_{\odot}$\,year$^{-1}$.

We note that the \ha-based value presented here only represents emission from the three circular apertures in Figure \ref{white}, while \citet{capa15} integrate over the full field. Integrating over the full field results in a $SFR_{\rm H\alpha}$ value that is $<1\sigma$ discrepant (see Appendix \ref{ISA}), suggesting that the star formation in this system is concentrated in the three line-emitting regions.

We may use the \siia/\siib ratio of each component to constrain the electron density (\citealt{prox14}). While the resulting uncertainties are large (i.e., up to $\sim1\,$dex), the broad contributions of all three components feature similar densities ($\sim10^3$\,cm$^{-3}$). Since no significant narrow \siia emission is detected from any component, we may not constrain the electron density of the source of the narrow emission. The $n_e$ value of the broad emission in each component is comparable to those of other $z\sim4-6$ galaxies (e.g., \citealt{isob23}). 

Our dust-corrected line ratios (see Table \ref{lineRtable}) are combined with the metallicity diagnostics of \citet{curt20} to determine the gas-phase metallicity of each component. Note that we only include line fluxes determined from the R2700 data (i.e., excluding \oiilow), as the complex line shapes and spectral blending cause R100-based line fluxes to be incorrectly estimated. Assuming a solar oxygen abundance of $\rm 12+\log_{10}(O/H)=8.69\pm0.05$ \citep{aspl09}, we find gas-phase metallicities of $\sim0.53-0.72$ solar for the three components (in agreement with \citealt{mark22}). These are higher than most galaxies in this epoch (e.g., \citealt{curt24}), but comparable to the metallicities found in other high-redshift merging systems (e.g., \citealt{arri23,rodr24,pere24,vent24}).

\begin{table}
\centering
\begin{tabular}{c|ccc}
 & HZ10-E & HZ10-C & HZ10-W \\ \hline
$E(B-V)_{\rm N}$ & $0.23\pm0.07$ & $-0.05\pm0.13$ & $0.81\pm0.23$\\
$E(B-V)_{\rm B}$ & $0.34\pm0.08$ & $0.67\pm0.14$ & $0.52\pm0.04$\\
$E(B-V)$ & $0.29\pm0.05$ & $0.41\pm0.09$ & $0.56\pm0.05$\\  \hline
$SFR_{\rm H\alpha,N}$ [M$_{\odot}$\,year$^{-1}$] & $9\pm2$ & $3\pm1$ & $30\pm22$\\
$SFR_{\rm H\alpha,B}$ [M$_{\odot}$\,year$^{-1}$] & $17\pm4$ & $66\pm29$ & $62\pm9$\\
$SFR_{\rm H\alpha}$ [M$_{\odot}$\,year$^{-1}$] & $25\pm4$ & $41\pm12$ & $84\pm13$\\  \hline
$\log_{10}(n_{\rm e,N}/cm^{-3})$ & $-$ & $-$ & $-$\\
$\log_{10}(n_{\rm e,B}/cm^{-3})$ & $3.02\pm0.95$ & $2.82\pm0.36$ & $3.25\pm1.18$\\  \hline
12+log$_{10}$(O/H)$_{\rm N}$ & $8.32\pm0.05$ & $8.50\pm0.05$ & $8.62\pm0.05$\\
$Z_{\rm N}$ [$Z_{\odot}$] & $0.43\pm0.07$ & $0.64\pm0.11$ & $0.85\pm0.14$\\
12+log$_{10}$(O/H)$_{\rm B}$ & $8.51\pm0.05$ & $8.54\pm0.05$ & $8.54\pm0.05$\\
$Z_{\rm B}$ [$Z_{\odot}$] & $0.66\pm0.11$ & $0.71\pm0.12$ & $0.70\pm0.11$\\
12+log$_{10}$(O/H) & $8.41\pm0.05$ & $8.49\pm0.05$ & $8.54\pm0.05$\\
$Z$ [$Z_{\odot}$] & $0.52\pm0.09$ & $0.63\pm0.10$ & $0.71\pm0.12$\\
\end{tabular}
\caption{Properties of HZ10 components derived from line ratio diagnostics. Included are the observed B-V colour excess derived from the Balmer decrement, the dust-corrected SFR derived from \ha intensity, the electron density derived from \siia/\siib ratios, and metallicities calculated using the diagnostics of \citet{curt20} and the line ratios listed in Table \ref{lineRtable}. We assume a \citet{salp55} IMF and \citet{calz00} extinction law.}
\label{ratvals}
\end{table}

\subsection{Spatially resolved analyses}\label{RAS}

While the component-based analysis allows us to characterise the three primary line-emitting regions and constrain some physical parameters, the nature of the NIRSpec IFU also allows us to explore these properties in a resolved manner. That is, we may extract spectra from each spaxel of the R100 and R2700 cubes and fit each to reveal the ionised gas kinematics, metallicity, and morphology of the complex source HZ10.

\subsubsection{R2700 resolved analysis}\label{r2700spx}
The R2700 data cube is a rich treasure trove of morpho-kinematic information, but the fact that the spectra of many spaxels exhibit non-symmetric emission lines complicates resolved analysis. These skewed lines could represent tidal interaction features or outflows (we examine evidence for the presence of a broad line region [BLR] in Section \ref{blrsec}), as well as contamination from emission in other spaxels within the PSF. 
We explore the R2700 data cube by fitting the spectrum of each spaxel using the flexible line and continuum model detailed in Section \ref{CA}.

The approach of fitting two components to each line (i.e., narrow and broad) is more suitable than a single-Gaussian approach, but may still simplify the distribution and source of flux in the HZ10 field. To test this, we performed a non-parametric kinematic analysis of the spectrally distinct \oiiib emission (Appendix \ref{r2700mom}). Since the results of the two-Gaussian and non-parametric approaches agree, the two-Gaussian approach is sufficient to capture the asymmetric morpho-kinematics of this field.

\begin{figure*}
\centering
\includegraphics[width=\textwidth]{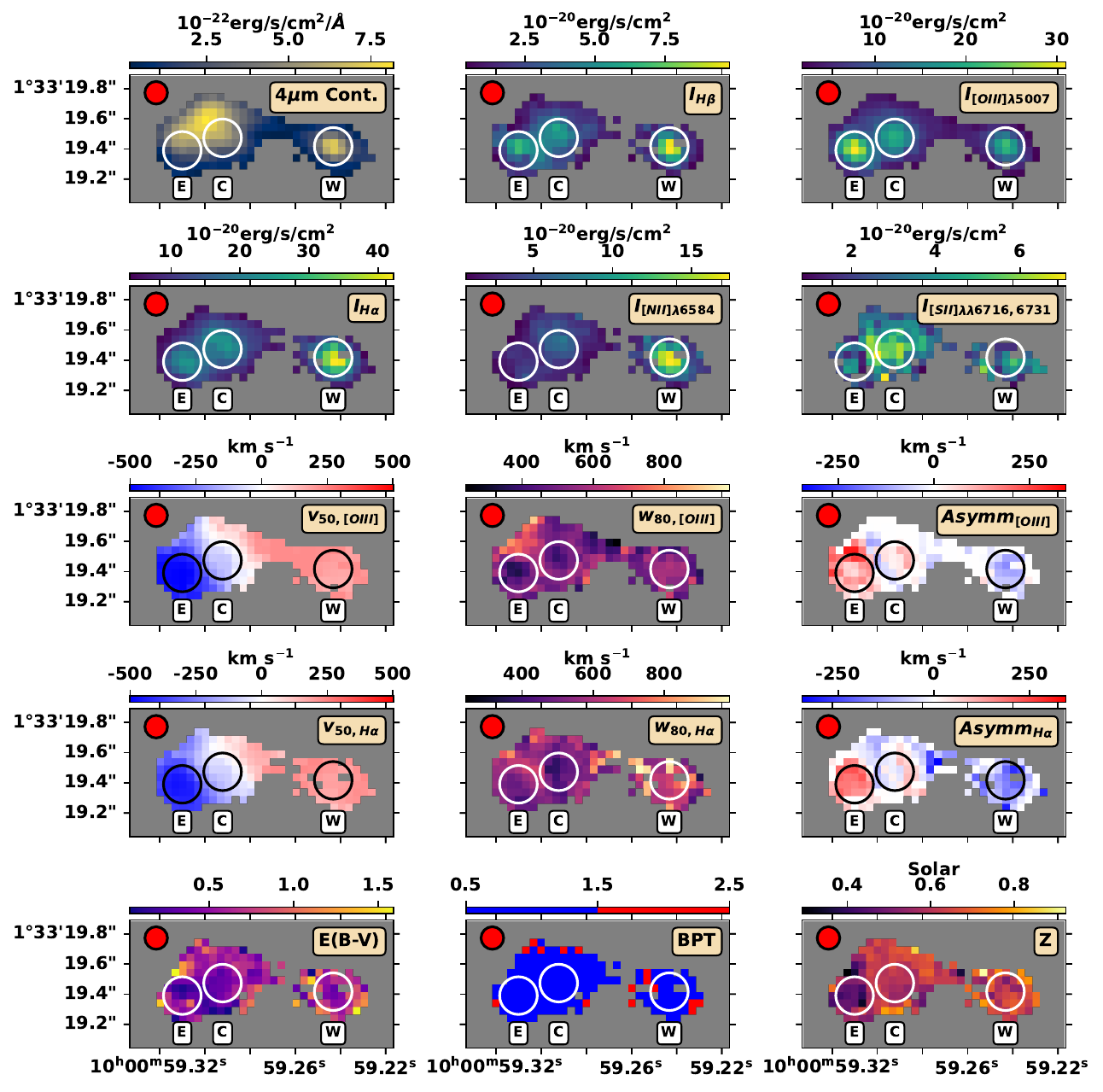}
\caption{Results of fitting the spectrum of each spaxel in the R2700 data cube. We include the continuum flux at $\lambda_{\rm obs}=4\,\mu$m, integrated line intensities, and kinematic measures. In addition, we present the colour excess E(B-V), the [NII]-BPT classification of \citet{schol23}, and the gas-phase metallicity. Hollow circles depict the location of each line peak (see Figure \ref{white}), while the fiducial PSF (FWHM$=0.15''$) is depicted by a red circle to the upper left.}
\label{r2700mom_real}
\end{figure*}

Using LMFIT in `least\_squares' mode, we first fit each spectrum with all variables free. The results are inspected, we fix the flux of all line components with S/N$<3$ to 0, and re-run the fit until only well-fit lines remain. Once the fit converges, we extract the continuum value, total line flux of each line, and velocity information from each best-fit model. The results are shown in Figure \ref{r2700mom_real}.

The fits are used to extract relevant kinematic measures for \oiiib and \ha. First, we consider $v_{\rm X}$, or the velocity (with respect to a systemic redshift of $z=5.6543$) at which the cumulative distribution function (CDF) of the best-fit line model (including both broad and narrow components) reaches a value of $X\%$. This results in estimates of $v_{50}$ , $w_{80}\equiv v_{90}-v_{10}$, and asymmetry$=|v_{50}-v_{90}|-|v_{50}-v_{10}|$ (e.g., \citealt{herv23}). These measures respectively represent the line-of-sight velocity, linewidth, and whether the line is skewed to the red or blue.

For each spaxel where \hb, \oiiib, \niib, and \ha are well detected, we apply the [NII]-BPT demarcation line of \citet{schol23} to test for extreme AGN-excitation. E(B-V) is calculated using the observed Balmer decrement for each spaxel when both \ha and \hb are well determined (i.e., $>1\sigma$). Dust-corrected line fluxes (see Table \ref{lineRtable}) are also used to determine the metallicity (\citealt{curt20}).

We find that the continuum emission peaks to the north of the field (Figure \ref{r2700mom_real}). Most of the line flux is concentrated in the three components, with the exception of the weaker \siiab, which is strongest between \hza and \hzb. Due to the faintness of \siiab, we are not able to present maps of $n_e$. The kinematics of \oiiiab and the other lines are comparable: a strong east-west velocity gradient with no major trends in $w_{80}$ (expect for a slightly wider linewidth in \hzc). \hza and \hzc feature red and blue asymmetry, respectively, while \hzb straddles a separate gradient of asymmetry. 

The distribution of properties is in agreement with the aperture-based analysis of Section \ref{LRA} (see Table \ref{ratvals}) and the total integrated analysis of Appendix \ref{ISA}. As seen in the first two rows of Figure \ref{r2700mom_real}, \hb, \ha, and \niib are brightest in \hzc, while \oiiib is brightest in \hza. This distribution results in a lower value of R3 for \hzc, and thus a higher metallicity (\citealt{curt20}). Most of the spaxels are SF-excited, with the possible exception of a few low-S/N pixels east of \hzc. \hzc shows the highest $E(B-V)$, followed by \hzb and then \hza. \hza shows a low metallicity ($\sim0.3-0.5$ solar), while \hzb and \hzc are higher ($\sim0.5-0.7$ solar).

\subsubsection{R100 resolved analysis}

Next, we may follow the same spectral fitting procedure outlined in Section \ref{CA} to fit the line and continuum emission in spectra extracted from each spaxel of the R100 data cube. The spectrum of each spaxel is extracted from the R100 cube and is fit using LMFIT in `leastsq' mode. Because these data feature a broad LSF ($\sim10^4$\,km\,s$^{-1}$), they are more suited for characterising continuum emission than line emission. Thus, we focus on the continuum emission, and only include line emission in our fits in order to reduce their bias on our fits. 

\begin{figure}
\centering
\includegraphics[width=0.5\textwidth]{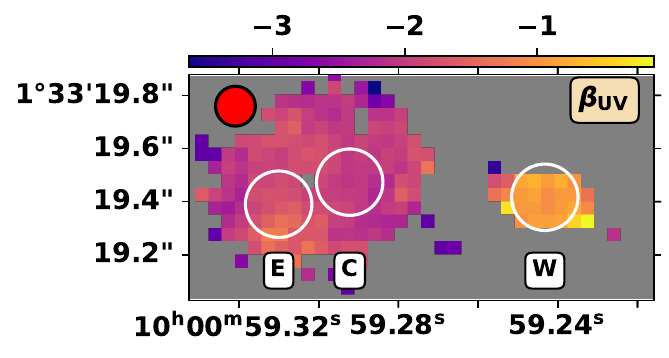}
\caption{The rest-UV power law slope ($\beta_{\rm UV}$), as derived by fitting the spectrum of each spaxel in the R100 data cube. Hollow circles depict the location of each line peak (see Figure \ref{white}), while the fiducial PSF (FWHM$=0.15''$) is depicted by a red circle to the upper left. The field of view is the same as in Figure \ref{r2700mom_real}.}
\label{r100_mom}
\end{figure}

The best-fit continuum power law slopes bluewards of the Balmer break ($\beta_{\rm UV}$) are recorded if the corresponding power law normalisations are well determined (i.e., $>1\sigma$). The resulting map (Figure \ref{r100_mom}) reveals that \hzc features a very red continuum $\beta_{\rm UV}\sim-1$, while the rest of the field features more standard values of $\beta_{\rm UV}\sim-2$. The values of $\beta_{\rm UV}$ in each component are further explored in Section \ref{natco}.

\section{Discussion}\label{discussion}

\subsection{Lack of evidence for presence of AGN}\label{agnbigsec}
From the previous Section, it is apparent that HZ10 is a complex system, with a strong velocity gradient and asymmetric emission lines. The emission line ratio diagnostics are not able to confirm or refute the presence of an AGN, in agreement with the analysis of \citet{satu18}. In this other work, the authors compared the location of AGN and SFGs in IRX-$\beta$ diagrams, finding a low probability for AGN domination in HZ10 (i.e., $\le 30\%$). Our JWST/NIRSpec data allow us to use other methods to search for AGN signatures. Here, we explore whether there is a broad line region (BLR) in this system (Section \ref{blrsec}), and if there are high-ionisation lines detected (Section \ref{agnsec}). 

\subsubsection{Testing for the presence of a BLR}\label{blrsec}

BLRs, which represent high-dispersion gas in the immediate surroundings of an accreting black hole, have been discovered in multiple $z>4$ galaxies with JWST/NIRSpec through the detection of broad (FWHM$>10^3$\,km\,s$^{-1}$) components of permitted lines, typically Balmer hydrogen lines (\ha and \hb; e.g., \citealt{maio23a,parl23,uble23b,loia24}). Crucially, forbidden lines (including \oiiiab, \niiab, and \siiab) are not emitted from BLRs due to suppression in high-density environments (e.g., \citealt{bask05,cres23}).

In Section \ref{CA}, we extracted spectra from the R2700 data cube using circular apertures of radius $0.125''$ centred on each of the three line peaks, and fit each with a two-component (broad and narrow) model for each line. To allow for the presence of a BLR, we allowed the linewidths of each forbidden line to differ from the linewidths of the Balmer lines. The resulting fits (Figure \ref{CA_2700}) and best-fit values (Table \ref{linetable_CA}) show that the broad component is redshifted in \hza and blueshifted in \hzc. Both the Balmer lines and \oiiiab feature a broad component with FWHM$<10^3$\,km\,s$^{-1}$. Because there is no strong evidence for an extreme broad component (e.g., \citealt{parl23,uble23b}), it is likely that the broad component in these spectra does not originate from an extreme BLR, and instead represents a tidal feature or outflow. However, we note that a weak BLR may still be present, contributing to the broad emission.

\subsubsection{Lack of high-ionisation lines}\label{agnsec}

We also conduct a search for high-ionisation lines (ionisation energies of $\sim50-100$\,eV): \civ, \heiia, \neiv, and \heiib (e.g., \citealt{felt16,lapo17,naka18,tozz23,maio23a}). A detection of these lines indicates a hard ionising spectrum (e.g., \citealt{topp24a}), and they have been used to construct line ratio diagnostics that separate AGN and star-forming environments more successfully than the [NII]-BPT and [SII]-VO87 diagrams (e.g., \citealt{shir12}). On the other hand, some works have used non-detections to argue against photoionisation by an AGN (e.g., \citealt{will14}). 

\begin{figure*}
\centering
\includegraphics[width=\textwidth]{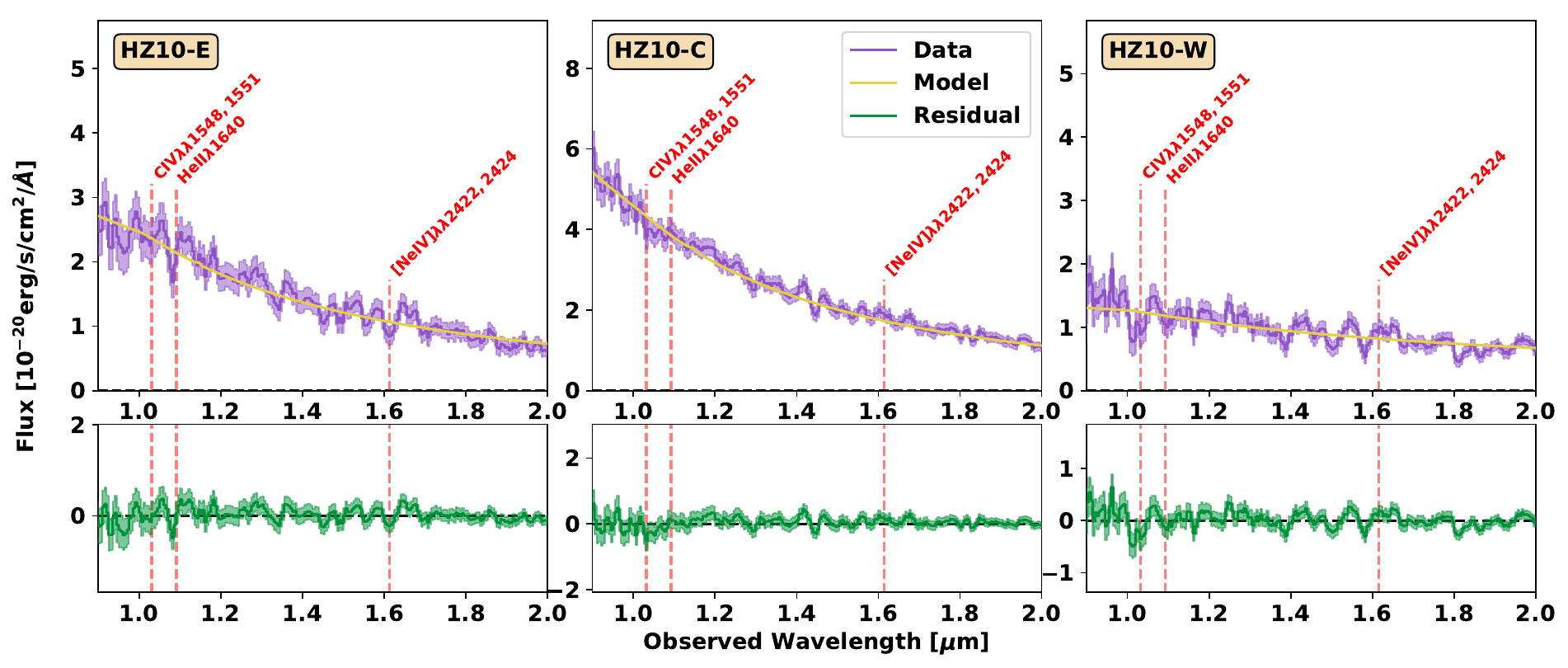}
\caption{Spectra extracted from the R100 data cube (purple lines), using circular apertures centred on each component, focused on three high-excitation emission lines (red vertical dashed lines). We present a fit to the continuum of each spectrum (yellow line), as well as residuals (green line). The $1\sigma$ RMS noise level is shown by the shaded regions.}
\label{rarelines}
\end{figure*}

We search for evidence of these lines in spectra extracted from the R100 data cube using the same component-focused apertures of Section \ref{blrsec}. The resulting spectra around \civ, \heiia, and \neiv are shown in Figure \ref{rarelines}, where we also present a fit to the continuum. It is clear that none are detected at high significance. Our data show no evidence for significant detections of the rest-optical line \heiib.

These non-detections suggest that the three primary line-emitting components of HZ10 are not dominated by high-ionisation regions. However, we note that the spectral range required for the detection of most of these lines is only covered by the low spectral resolution R100 data. Higher spectral resolution observations may reveal fainter and narrower lines (e.g., \citealt{bunk23,maio23a}). Through observations of 25 quasars at $z>6.5$ with JWST/NIRCam (`A SPectroscopic survey of biased halos In the Reionization Era'; ASPIRE; \citealt{wang23}), \citealt{yang23} only found evidence for significant \heiib emission from one source, suggesting that the non-detection of this line does not exclude the possibility of AGN activity.

\subsection{Comparison with ALMA \cii Data}
The JWST data presented in this work allow us to characterise the UV- and optical-bright emission of HZ10, but we are still plagued by the effects of dust extinction. This may be examined through the Balmer decrement and colour excess E(B-V) (see Figures \ref{r2700mom_real} and \ref{r100_mom}), but it is still possible that there are highly attenuated areas whose faint UV emission falls beneath our sensitivity limit. To study areas possibly affected by strong dust obscuration, we leverage high-spatial resolution ALMA data targeting \cii emission in this field from the CRISTAL survey \citep{herr25}, which are analysed in a parallel work (\citealt{teli24}). 

Many works have shown that \cii is a suitable tracer of the total SFR of galaxies in the local and $z>4$ Universe (e.g., \citealt{delo14,scha20}). However, recent works have argued that it is better suited as a tracer of the atomic (e.g., \citealt{vizg22b}) or molecular gas (e.g., \citealt{zane18,dess20,vizg22a,deug23b,arav24}), and observations of low metallicity ($Z\sim0.2-0.7$ solar) dwarf galaxies suggest that \cii may be a better tracer of total molecular gas content than the commonly used CO or [CI] tracers (\citealt{madd20}). This complexity arises partly from the low ionisation potential of carbon (11.26\,eV), which enables \cii emission from multiple phases of the ISM (i.e., warm and dense molecular gas, warm or cold neutral gas; e.g., \citealt{lang10,pine13,gurm24}). In this work, we interpret \cii as a tracer of potential and/or ongoing star formation that is less affected by dust obscuration than rest-UV or rest-optical tracers.

\begin{figure*}
\centering
\includegraphics[width=\textwidth]{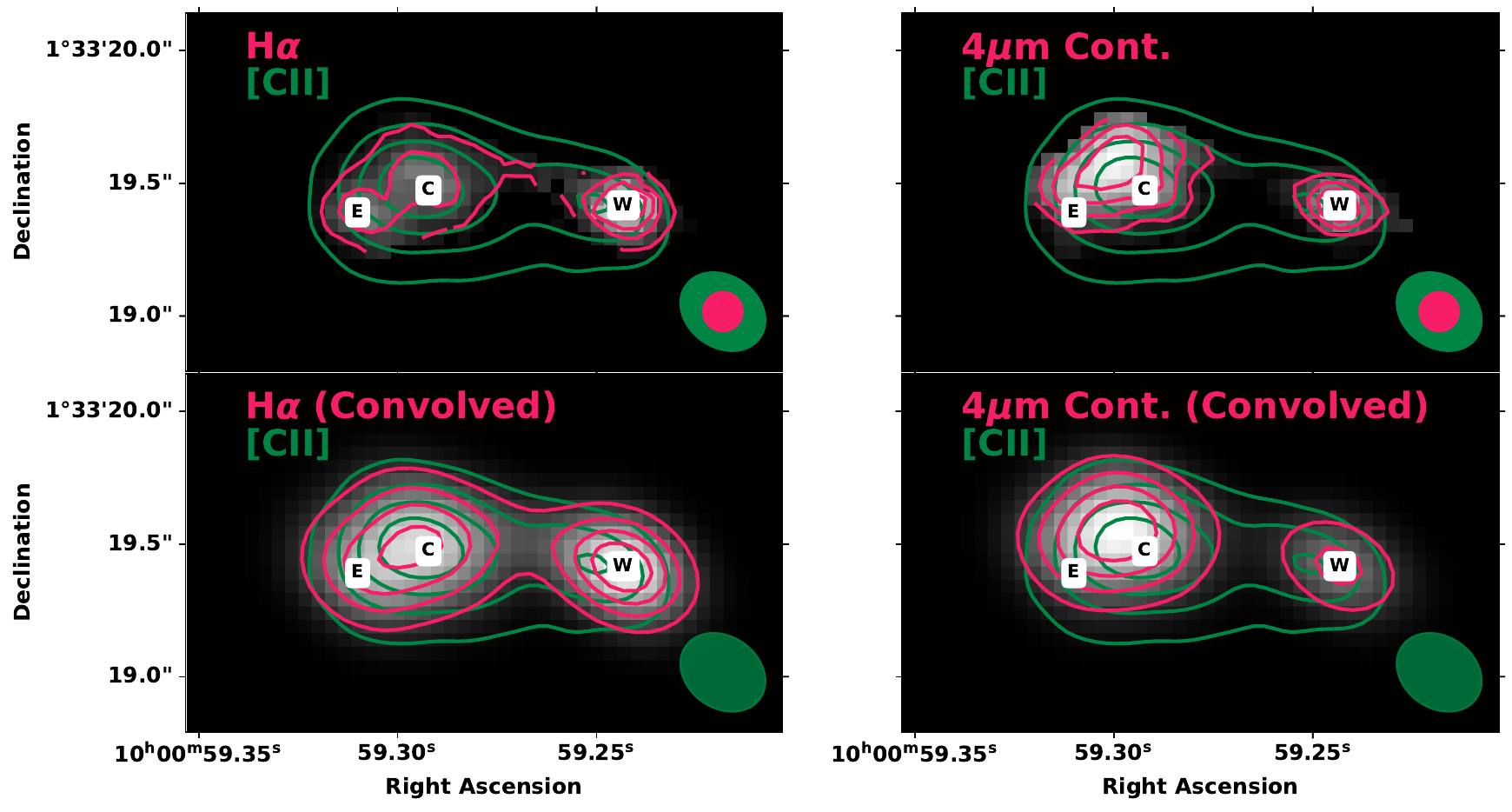}
\caption{Comparison of H$\alpha$ (pink contours and greyscale in left column) and $\lambda_{\rm obs}=4\,\mu$m emission (pink contours and greyscale in right column) from our JWST/NIRSpec R2700 data cube (Section \ref{r2700spx}) with ALMA \cii emission (green contours in each panel; \citealt{teli24}). Contours are shown at [20, 40, 60, 80]$\%$ of the maximum value of each map. The top row shows the NIRSpec data at their native resolution, while NIRSpec data in the lower row are convolved to the same resolution as the ALMA data. The approximate PSFs are shown by filled ellipses.}
\label{ciiuv}
\end{figure*}

We compare the ALMA \cii distribution with the observed \ha and $\lambda_{\rm obs}=4\,\mu$m continuum maps derived through spaxel-by-spaxel fits to the R2700 cube (see Section \ref{r2700spx}) in Figure \ref{ciiuv}. The astrometry of both datasets have been aligned to the Gaia DR3 system, so no further spatial shifts are required. The restoring beam of the ALMA data (FWHM of $0.343''\times0.271''$ at a position angle of $54.6^{\circ}$) is large relative to the NIRSpec IFU PSF (FWHM$\sim0.1-0.2''$, \citealt{deug23a}; we assume a fiducial FWHM$=0.15''$), so a direct comparison is slightly ambiguous (top row of Figure \ref{ciiuv}). To better compare these data, we convolve the NIRSpec dataset with a custom Gaussian kernel to match the ALMA PSF (photutils function create\_matching\_kernel; \citealt{brad22}), as shown in the bottom row of Figure \ref{ciiuv}. 

By design, the \ha emission (tracing unobscured star formation) is focused in the three line-emitting regions. When convolved to match the ALMA beam to compare to the \cii morphology, we find a slight east-west offset between the two tracers for \hzc and a good match for \hzb (as found by \citealt{teli24}). The \cii emission does not feature an extension towards \hza (as seen in \ha), suggesting the lack of a strong \cii peak in \hza. While the $\lambda_{\rm rest}=4\,\mu$m emission shows a similar morphology as the \ha emission for \hzc (i.e., an east-west offset with respect to the \cii emission), we find that the $\lambda_{\rm rest}=4\,\mu$m emission is offset to the northeast from \hzb.

Offsets between \cii and UV emission are commonly observed at high redshift (e.g., \citealt{maio15,inou16,carn17,carn18,kill24}). It has been suggested that these offsets may be caused by a non-isotropic dust distribution, as \cii may be observed from dusty regions of the cold neutral medium (e.g., \citealt{katz17}). Indeed, \citealt{teli24} find that the $\lambda_{\rm rest}\sim158\,\mu$m FIR continuum emission shows a very similar morphology to the \cii emission, with the exception of a possible `bridge' between \hzb and \hzc (as found in \citealt{vill24}). Since this dust would strongly attenuate \ha, it is possible that the east-west offset between \cii and \ha that we see in \hzc represents areas of obscured and unobscured star formation (as supported by the red $\beta_{\rm UV}>-1$ of \hzc, see Table \ref{ratvals}). The offset of the $\lambda_{\rm rest}=4\,\mu$m emission from \hzb may indicate a region of low ongoing star formation but high stellar mass.

\subsection{Nature of components}\label{natco}

In this work, we have used the R100 and R2700 data cubes to characterise the HZ10 system in both a component-based and spatially resolved manner. The line emission is concentrated in three peaks (here called \hza, \hzb, and \hzc), while the continuum emission overlaps these peaks and features additional emission to the north of \hzb. All three components show broad \oiiiab and Balmer lines, which imply tidal features or outflows rather than the presence of a BLR. We cannot confirm the presence of an AGN in each component, as common high-ionisation lines are not detected. Here, we synthesise our analyses to propose the natures of different parts of this field.

\subsubsection{\hzc}\label{hzcsec}

Because \hzc is spatially separated from the other emission, features overlapping line and continuum emission, and has a nearly uniform $w_{80}$, \oiiib rest-frame equivalent width (see Appendix \ref{r2700mom}), and metallicity, we conclude that this is an independent galaxy with a coincident stellar and gas component. Its small spatial separation ($<5$\,kpc, projected) and line of sight velocity offset ($<500$\,km\,s$^{-1}$) suggest that it is kinematically associated with the rest of HZ10. It features a continuous velocity gradient with \hzb, which was previously interpreted as evidence for a single rotating system (e.g., \citealt{jone17}). However, such velocity gradients are often observed in merging systems, where they could represent tidal interactions (e.g., \citealt{smit18,hash19,bakx20,scho24}). There is little to no velocity gradient across the area of \hzc, but it features a higher velocity dispersion compared to \hza and \hzb (as explored through $w_{80}$) and strong blue asymmetry in all lines. This combination could imply past tidal interactions or an outflow.

It is clear that \hzc has a red continuum ($\beta_{\rm UV}\gtrsim-1$). An in-depth analysis of JWST/NIRCam data from the JWST Advanced Deep Extragalactic Survey (JADES) revealed that the median UV spectral slope for galaxies at $z\sim5.9$ is $\beta_{\rm UV}=-2.26\pm0.03$ \citep{topp24b}. While values around $-2$ are commonly observed (e.g., \citealt{dunl13,bouw14}), there have been cases of very red objects at high redshift (e.g., $\beta_{\rm UV}\sim-1.2$ at $z\sim7$ in \citealt{smit18}; $\beta_{\rm UV}\sim-1.1$ at $z\sim6.5$, \citealt{matt19}). At a given redshift, $\beta_{\rm UV}$ has been found to correlate with $M_*$, and anti-correlate with M$_{\rm UV}$ (e.g., \citealt{bhat21}). However, the intrinsic and observed UV slopes may differ due to dust attenuation (e.g., \citealt{wilk13}), so redder slopes may indicate larger amounts of dust. This is supported by the east-west offset observed between the FIR continuum and \ha, which may indicate an obscuring dust reservoir. Regardless, \hzc features the reddest continuum slope, highest metallicity, and highest E(B-V) of the three HZ10 line-emitting components (Table \ref{ratvals}), implying a more enriched system than the rest of HZ10.

\subsubsection{\hza and \hzb}

\hza and \hzb are more complex, as they feature a velocity gradient, independent areas of low $w_{80}$, and separate peaks of line emission. If they represented a single regular rotating disk, then we would expect to see coincident peaks of line and continuum emission in the centre of a symmetric velocity gradient. Since the peaks are offset from the centre, this could instead represents a pair of galaxies. We note that two distinct peaks could also arise in a disc with a more dust-obscured kinematic centre. While we find little evidence of high obscuration in the spatial region between the two emission peaks, this is still a possibility. \hza also exhibits a redder $\beta_{\rm UV}$, higher E(B-V), and lower metallicity than \hzb, arguing in favour of two distinct galaxies. 

From comparison between the line and continuum emission (Fig. \ref{white}), it is possible that there may be additional galaxies near \hza and \hzb. The region to the northeast of these line peaks is UV-luminous but faint in line emission, suggesting a gas-poor population of stars that may be seen immediately post-starburst ($\sim10$\,Myr). Further evidence for this is given by a comparison of the JWST data to new ALMA \cii data, which shows that this northeast region is not \cii-luminous. It is true that this region does not show drastic signs of difference in the gas-phase diagnostics (e.g., $\beta_{\rm UV}$, $Z$, E(B-V)), but this may signify that these diagnostics are dominated by gas stripped from \hza and \hzb. This also argues against the idea that the northeast emission represents a low-redshift interloper.

\subsubsection{Summary}

Together, these findings suggest that the eastern emission in HZ10 is composed of at least three primary constituents: two star-forming galaxies featuring broad components redshifted with respect to their narrow emission (\hza and \hzb) and a gas-poor stellar component to the north. The western emission (\hzc) features a blueshifted broad component (with respect to its narrow emission) and coincident UV and line emission. Due to the proximity of these sources, this complex picture may suggest an ongoing merger of multiple galaxies. 

In order to classify `close' pairs of galaxies, it is common to apply the criteria of $\Delta r \lesssim20$\,kpc and $\Delta$v$\lesssim500$\,km\,s$^{-1}$ (e.g. \citealt{dunc19,vent19,roma21}). All of the components in HZ10 meet the spatial separation criterion ($<5$ projected kpc), and the neighbouring components (i.e., \hza /\hzb and \hzb /\hzc) meet the line-of-sight velocity criterion. Thus, it is likely that all three line-emitting galaxies will merge.

Simulations have shown that major mergers may result in multiple clumps of gas within $\sim5$\,kpc (e.g., \citealt{koha19,rizz22}), while observations at cosmic noon ($z\sim2$) have revealed that close mergers may feature nearly continuous velocity gradients (e.g., \citealt{simo19}), as is the case here. The presence of such gas-rich and gas-poor components in the same system is odd at first glance, but similar systems have already been discovered at high redshift (e.g., Jekyll and Hyde; \citealt{schr18,pere24}). 

As discussed in \citet{teli24}, this interpretation is just one possibility (see their figure 8). The eastern emission (containing \hza and \hzb) may represent a single disturbed disk, a merger of two galaxies, or a merger of three galaxies. The single-galaxy possibility is supported by the regular velocity field, but the asymmetry pattern and morphology of line and continuum emission imply complex kinematics and significant variations in ISM conditions. If this is instead a two-galaxy merger, then the line peaks would represent each galaxy, and the asymmetry pattern could be caused by tidal interactions. Finally, \hzb may be composed of two galaxies undergoing a very close-scale merger. This system shows a similar morphology to the $z=4.56$ system DC818760, which was originally detected as a triple merger \citep{jone20}, but was resolved into at least four galaxies through higher-resolution observations \citep{deve20}. In order to determine the true nature of \hza and \hzb, additional high-resolution observations are required (e.g., JWST/NIRCam).

\subsection{Comparison to HZ6 \& AzTEC-3}
Based on photometric data, \citet{pave18} find that HZ10 and the starburst galaxy CRLE lie within a galaxy overdensity at $z\sim5.7$. This situation is quite similar to that of the starburst galaxy AzTEC-3 and star-forming galaxy HZ6\footnote{HZ6 is also named 
LBG-1 (e.g., \citealt{riec14}), DEIMOS\_COSMOS\_848185 (e.g., \citealt{lefe20}), and CRISTAL-02 (e.g., \citealt{iked25}).} at $z=5.3$ (\citealt{capa11,riec14}). Both systems feature a starbursting galaxy ($SFR\sim10^3$\,M\,$_{\odot}$\,yr$^{-1}$) with a more representative, star-forming galaxy within a projected distance of $100$\,kpc and a comparable overdensity within a radius of 2\,cMpc. 

Furthermore, the morpho-kinematics of HZ6 and HZ10 both indicate complex merger behaviour on kpc-scales. A detailed analysis of high-resolution [CII] data from the CRISTAL survey and archival HST/WFC3 F160W images \citep{iked25} reveals that the [CII] morphology of HZ6 is well-described by a single 2D Gaussian model, while the rest-UV emission is resolved in three clumps. This is the only CRISTAL source that shows a broad wing in the [CII] spectrum (which may indicate merging or outflow activity; Davies et al. in prep), which was already seen in lower-resolution data from ALPINE \citep{beth20}. While results of JWST/NIRSpec IFU observations of HZ6 (PID 3045, PI A. Faisst) are not yet available, the picture from current data is that HZ6 represents a close-separation merger of at least three galaxies (\citealt{pave19,fais20,iked25}). With the current JWST and ALMA data, it is clear that HZ10 is a merger of at least two galaxies (see Section \ref{natco} and discussion of \citealt{teli24}). 

While wide-field observations around FIR-bright galaxies have resulted in the detection of galaxy overdensities (e.g., \citealt{bisc18,pave18,gino22,calv23,zewd24}), more detailed analyses of the galaxies that are not the brightest members are needed. Our current data allow us to characterise the morpho-kinematics and ISM conditions of such galaxies on kpc-scales. HZ10 and HZ6 are both complex systems, with multiple closely associated components ($<5$\,projected kpc). Our analysis of HZ10 shows that one component (\hzc) is more enriched than the others, possibly indicating a different star formation history. Additional focused observations of non-dominant galaxies in high-redshift overdensities will enable characterisation of their current properties, placing constraints on their histories and informing cosmological models and simulations (e.g., \citealt{hash23b,renn24}).

\section{Conclusions}\label{conclusion}

In this work, we present both high ($R\sim2700$) and low spectral resolution ($R\sim100$) JWST/NIRSpec IFU data of the $z\sim5.7$ dusty star-forming galaxy HZ10, as part of the GTO program GA-NIFS. The high spatial resolution (PSF$\sim0.15''$) of our data opens a new window into the stellar and ionised gas properties of this source, revealing multiple components within a $\sim1.5''$ diameter region: three components detected in emission lines (\hza, \hzb, and \hzc) and an additional offset rest-optical continuum source.

We extract spectra from the R2700 data cube using circular apertures centred on each line-emitting component and fit each with a combined continuum and line model (\hb, \oiiiab, \niiab, \ha, \siiab). Each line is well fit by a two-component model, suggesting the presence of tidal features. Through line ratio analyses, we find little evidence for AGN domination ([NII]-BPT and [SII]-VO87 diagrams), a high metallicity ($0.5-0.7$ solar, using the strong line diagnostics of \citealt{curt20}), and electron densities of the broad emission comparable to other galaxies at $z\sim4-6$: $\log_{10} ( n_e / cm^{-3} ) \sim3$. The R100 spectrum (which features a wider wavelength range) also exhibits emission from the unresolved doublet \oiilow, a range of UV spectral slopes ($\beta_{\rm UV}\sim-2.0$ to $-1.0$), and no significant \lya emission (with an upper limit in agreement with a previous detection, see Appendix \ref{ISA}).

The great potential of the IFU data is exploited by then exploring each cube on a spaxel-by-spaxel basis. Line emission is focused in the three primary components, while continuum emission ($\lambda_{\rm obs}=4\,\mu$m, or $\lambda_{\rm rest}\sim0.60\,\mu$m) features an extension to the north. A significant east-west velocity gradient is detected. \hza and \hzb show red line asymmetries, while \hzc features blue asymmetry. No major trends in metallicity or E(B-V) are detected, while \hzc is found to exhibit a red spectral slope $\beta_{\rm UV}\sim-1$.

We find that since all of the lines (i.e., both Balmer and forbidden lines) features both a narrow and broad component with $\rm FWHM<10^3$\,km\,s$^{-1}$, there is no evidence for a strong BLR. In the same regions, we do not significantly detect high-ionisation lines that would indicate AGN photoionisation. Thus, the asymmetric line profiles in the current data may indicate the presence of outflows, tidal tails, and/or minor satellites. Further observations at higher spectral resolution are required to confirm this.

Through a comparison to new high-resolution ALMA \cii data (\citealt{teli24}), we find that the \ha and \cii emission are nearly spatially aligned, indicating ongoing star formation from molecular gas reservoirs. Some offsets are detected, indicating dust reserviors that are obscuring rest-optical and rest-UV emission. On the other hand, the rest UV-optical continuum emission to the north of \hzb is offset from the \cii, suggesting a gas-poor population.

Together, our analyses suggest a re-interpretation of the well-studied source HZ10. While previous observations at low spatial resolution suggested a galaxy with a strong rotation gradient and offsets between some tracers, we are now able to resolve multiple separate galaxies. Due to their close spatial and spectral association, they likely represent a merging group. One galaxy is spatially separated, features nearly coincident rest-optical continuum, \cii, and line emission (\hzc), and has a blue line asymmetry. To the east of this object lies a complex group of two line emitting peaks (\hza and \hzb) with a strong velocity gradient and red line asymmetry. Furthermore, there is a UV-bright population of gas-poor stars to the north of these galaxies which may represent a separate galaxy or a distinct region of \hzb. The current data are not yet sufficient to state whether the emission east of \hzc represents a single disturbed disk or a multiple merger.

The analysis presented here represents a major step forward in the characterisation of the ionised gas morpho-kinematics, ISM conditions, and general nature of a FIR-bright, high-redshift group of galaxies. There are still some ambiguities (e.g., the distribution of \lya emission, the presence of narrow-line AGN), which may be remedied by higher spectral resolution NIRSpec/IFU observations in different gratings (e.g., G140H). Future SED analysis of the R100 data presented in this work will result in new estimates on stellar mass and star formation history. Work is already underway on ALMA high-resolution \cii (\citealt{teli24}) and FIR-emitting dust analyses \citep{vill24}. When complete, this field will act as a key prototypical close-separation merger at high-redshift for studies of galaxy evolution.

\section*{Acknowledgements}
We thank the anonymous referee for constructive feedback that enhanced this work.
GCJ and AJB acknowledge funding from the ``FirstGalaxies'' Advanced Grant from the European Research Council (ERC) under the European Union’s Horizon 2020 research and innovation programme (Grant agreement No. 789056).
GCJ, FDE and RM acknowledge support by the Science and Technology Facilities Council (STFC), by the ERC through Advanced Grant 695671 ``QUENCH'', and by the UKRI Frontier Research grant RISEandFALL.
KT was supported by ALMA-ANID grant number 31220026. KT, MA, and RHC acknowledge support from the ANID BASAL project FB210003.
SA, MP, and BRdP acknowledge grant PID2021-127718NB-I00 funded by the Spanish Ministry of Science and Innovation/State Agency of Research (MICIN/AEI/ 10.13039/501100011033).
SCa acknowledges support from the European Union (ERC, WINGS,101040227).
H\"U acknowledges funding by the European Union (ERC APEX, 101164796). Views and opinions expressed are however those of the authors only and do not necessarily reflect those of the European Union or the European Research Council Executive Agency. Neither the European Union nor the granting authority can be held responsible for them.
GC acknowledges the support of the INAF Large Grant 2022 ``The metal circle: a new sharp view of the baryon cycle up to Cosmic Dawn with the latest generation IFU facilities''.
RHC thanks the Max Planck Society for support under the Partner Group project “The Baryon Cycle in Galaxies” between the Max Planck for Extraterrestrial Physics and the Universidad de Concepción.
IL acknowledges support from PID2022-140483NB-C22 funded by AEI 10.13039/501100011033 and BDC 20221289 funded by MCIN by the Recovery, Transformation and Resilience Plan from the Spanish State, by NextGenerationEU from the European Union through the Recovery and Resilience Facility, and from PRIN-MUR project “PROMETEUS”  financed by the European Union -  Next Generation EU, Mission 4 Component 1 CUP B53D23004750006.
PGP-G acknowledges support from grant PID2022-139567NB-I00 funded by Spanish Ministerio de Ciencia e Innovaci\'on MCIN/AEI/10.13039/501100011033, FEDER {\it Una manera de hacer Europa}.
VV acknowledges support from the ALMA-ANID Postdoctoral Fellowship under the award ASTRO21-0062.
GCJ would like to thank Elena Bertola, Cosimo Marconcini, and Aayush Saxena for useful discussion and feedback on the manuscript.
This paper makes use of the ALMA data: ADS/JAO. ALMA\#2019.1.01075.S. ALMA is a partnership of the ESO (representing its member states), NSF (USA) and NINS (Japan), together with NRC (Canada), MOST and ASIAA (Taiwan), and KASI (Republic of Korea), in cooperation with the Republic of Chile. The Joint ALMA Observatory is operated by the ESO, AUI/NRAO, and NAOJ.

\section*{Data Availability}

The NIRSpec data used in this work was been obtained within the NIRSpec-IFU GTO programme GA-NIFS (PID 1217) and is publicly available. Data presented in this work will be shared upon reasonable request to the corresponding author.



\bibliographystyle{mnras}
\bibliography{example} 





\appendix

\section{R100/R2700 astrometry}\label{astromapp}
\begin{figure*}
\centering
\includegraphics[width=\textwidth]{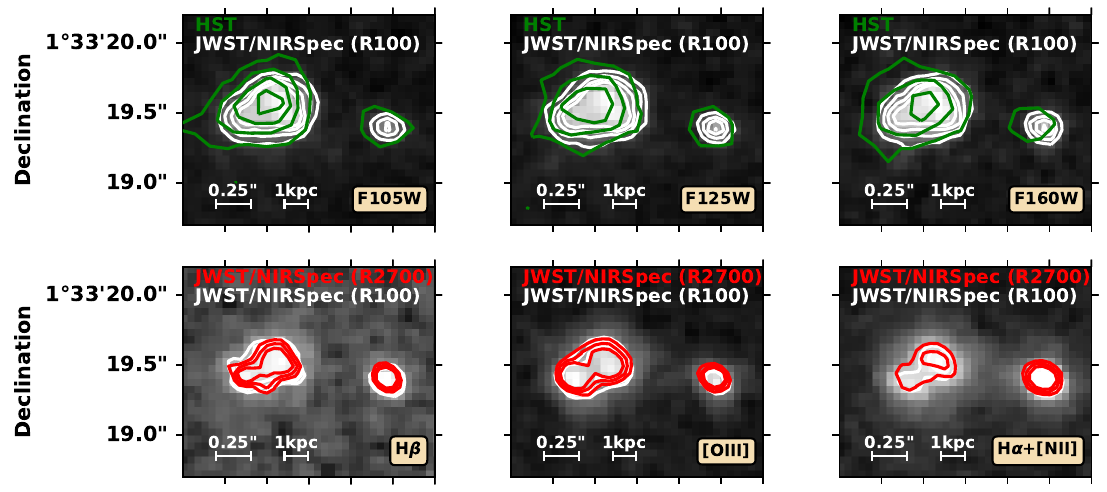}
\caption{Comparison of astrometric alignment of data using integrated intensity maps of JWST/NIRSpec IFU and HST data. The top row compares JWST/NIRSpec IFU (R100, background greyscale and white contours) and HST images (green contours) over three HST/WFC3 filters. The bottom row instead compares the two JWST/NIRSpec IFU data cubes (R100, background greyscale and white contours; R2700, red contours) over the wavelengths corresponding to three emission lines.}
\label{r100r2700}
\end{figure*}

As mentioned in Section \ref{obscal}, we verified that the R100 data was aligned with the Gaia DR3 astrometric frame through comparisons to aligned HST images from the MAST archive. These comparisons are shown in the top row of Figure \ref{r100r2700}. We also verify that the R100 and R2700 data cubes are aligned to the same frame by collapsing each over wavelength ranges corresponding to \hb, \oiiiab, and \ha+\niiab (see bottom row of Figure \ref{r100r2700}). As seen in Figure \ref{r100r2700}, these maps thus do not suffer from any astrometric issues (i.e., rotation, scaling, or systematic offsets).

\section{Integrated spectral analysis}\label{ISA}

In the main body of this work, we characterised the HZ10 field by fitting spectra extracted from the line peaks (Section \ref{CA}) and from each spaxel (Section \ref{RAS}). This revealed that HZ10 is not a single homogeneous source, but is composed of several galaxies with different conditions and kinematics. Since an analysis of the full field will average over all of the individual galaxies, we avoided this analysis. But since previous works examined HZ10 as a single system, it is worthwhile to study the integrated R100 spectrum for comparison to the literature. The kinematic complexity of HZ10 (i.e., asymmetric lines) dictates that this approach does not capture the full behaviour of the system. Thus, we only present selected results from this analysis.

\begin{figure*}
\centering
\includegraphics[width=\textwidth]{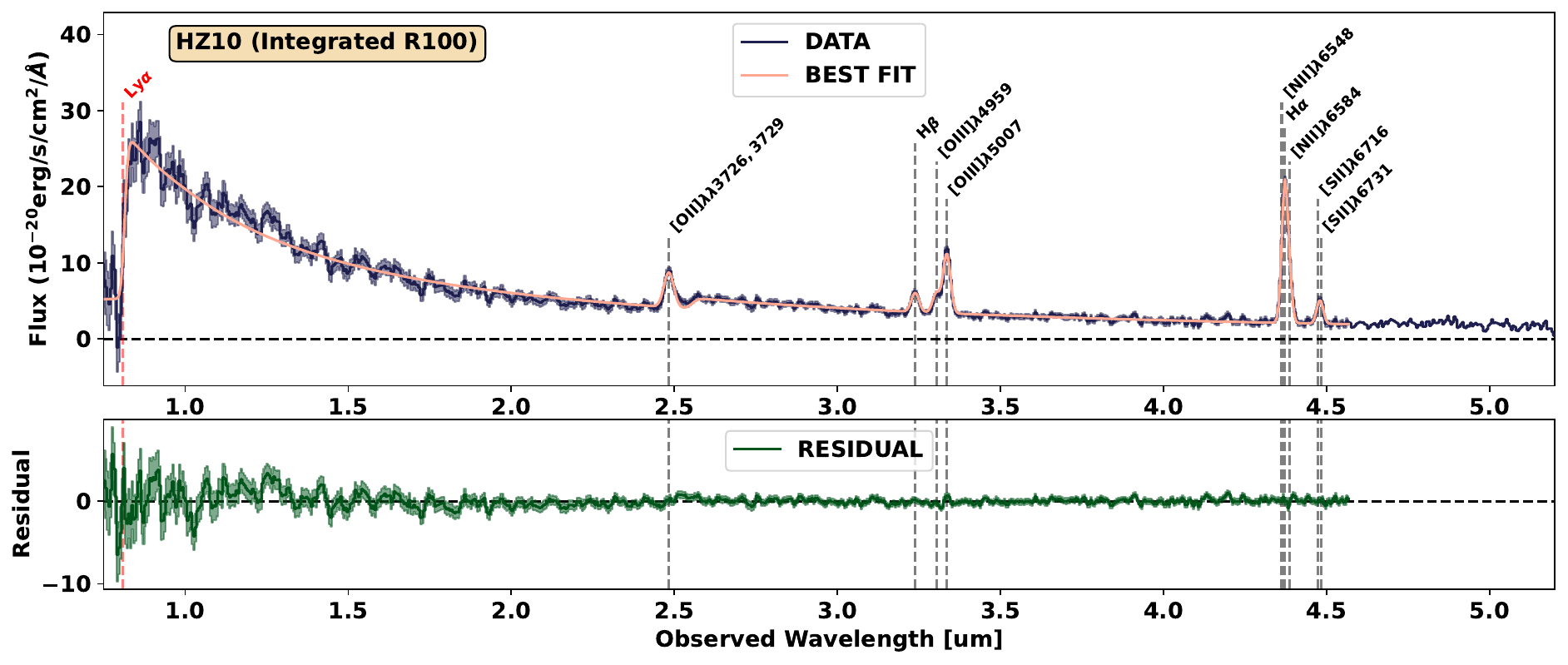}
\caption{Integrated spectrum extracted from the R100 data cube using the elliptical aperture of Figure \ref{white}. The best-fit model is shown by the tan line, as well as the residuals (green line). The wavelengths of each line (using the best-fit $z_{\rm sys}$) are shown by dashed vertical lines. The expected wavelength of \lya (which is not detected) is marked with a dashed red line. Uncertainties ($1\sigma$) for the extracted spectrum and residuals are shown by shaded regions.}

\label{whitespec}
\end{figure*}

We begin by extracting integrated spectra from the R100 data cube using a large elliptical aperture centred on 10m00m59.27s 01$^{\circ}$33$'$19.40$''$ with major and minor axes of $1.6''$ and $1.0''$, respectively (see dashed white ellipse in Figure \ref{white}). This aperture was chosen to contain all significant line and detection emission detected by our JWST/NIRSpec IFU observations, and is larger than the elliptical aperture used by \citet{teli24} to examine the total \cii spectrum. The line and continuum emission of the extracted spectrum is fit using a procedure identical to that of Section \ref{CA}, and the resulting fit is shown in Figure \ref{whitespec}

The continuum is well-detected in the R100 spectrum, resulting in $M_{\rm UV}=-22.22\pm0.02$ and $\beta_{\rm UV}=-1.71\pm0.02$, in agreement with previous observations (e.g., \citealt{capa15}). The best-fit Balmer line fluxes result in a colour excess that lies between the values of each component ($E(B-V)=0.44\pm0.22$). Correcting for this dust attenuation results in a large $SFR_{\rm H\alpha}=380\pm200$\,M$_{\odot}$\,yr$^{-1}$, where the large uncertainty is due to the large uncertainty in $E(B-V)$. This value is in agreement with both the $SFR_{\rm [CII]}=169_{-27}^{+32}$\,M$_{\odot}$\,yr$^{-1}$ from \citet{capa15} and the total $SFR_{\rm H\alpha}$ of the three line-emitting components we study here ($SFR_{\rm H\alpha}=150\pm18$\,M$_{\odot}$\,yr$^{-1}$). Similarly, the total best-fit metallicity ($Z=0.65\pm0.05$\,solar) is comparable to the metallicities of each line-emitting region (see Table \ref{ratvals}.)

We do not detect \lya emission in the R100 integrated spectrum, which yields a $3\sigma$ upper limit on the \lya rest-frame equivalent width of $REW_{\rm Ly\alpha}<11\angstrom$. \citet{mura07} detected \lya with an observed-frame equivalent width of $75\pm22\angstrom$, which corresponds to a rest-frame value of $ 11.3\pm3.3\angstrom$ (assuming $z_{\rm \cii}$ from \citealt{pave16}). Since this previous work predicted the \lya flux using a large circular aperture of diameter $3''$, this slightly larger $REW_{\rm Ly\alpha}$ may suggest that HZ10 exhibits a \lya halo (although an integration over the full IFU field of view does not return a significant \lya detection). On the other hand, the assumed redshift in the previous work ($z=5.7$) is slightly larger than the true redshift of this source, and the $REW_{\rm Ly\alpha}$ value is based on fits to photometry. Taking into account the uncertainties, our non-detection of \lya is not in disagreement with previous results, although follow-up observations at higher spectral resolution (e.g., JWST/NIRSpec IFU G140H/F070LP or the Multi Unit Spectroscopic Explorer [MUSE] on the Very Large Telescope [VLT]) are needed to confirm this.

\section{[OIII]5007 non-parametric kinematic analysis}\label{r2700mom}
Throughout this work, we have modeled each line in the R2700 spectra using two Gaussians (i.e., narrow and broad). However, this simple model may underestimate the complexity of the system. To examine this, we produce non-parametric morpho-kinematic maps for the strong, spectrally distinct line \oiiib and compare them to maps produced using our two-Gaussian model.

First, we isolate all data within $\pm1500$\,km\,s$^{-1}$ of the systemic redshift (here assumed to be equal to $z_{\rm [CII]}=5.6543$; \citealt{pave16}) and extract spectra for each spaxel. The continuum level for each spectrum is assumed to be constant, and is derived by creating a cumulative distribution function (CDF) of the line-free data (here assumed to be all data in [-1500,-750]$_{\cup}$[750,1500]\,km\,s$^{-1}$) and fitting this with a first-order polynomial (scipy curve\_fit). If the resulting fit features a positive, well-determined slope, it is used as the continuum flux for the spaxel, which is then subtracted from the spectrum.

The resulting line-only spectra for each spaxel are then used to derive spectral CDFs, which are normalised such that their highest and lowest values are set to unity and zero, respectively. Using \textlcsc{pygam}\footnote{A python package to fit generalized additive models (\url{https://pygam.readthedocs.io/en/latest/}.)} \citep{serv18}, we fit each normalised CDF with a monotonically increasing spline function. If the best fit spans the central $80\%$ of the continuum CDF (i.e., a minimum value less than 0.1 and a maximum value greater than 0.9) and \textlcsc{pygam} reports a good fit (i.e., pseudo-R-squared value greater than 0.9), we record the best-fit model.

\begin{figure*}
\centering
\includegraphics[width=\textwidth]{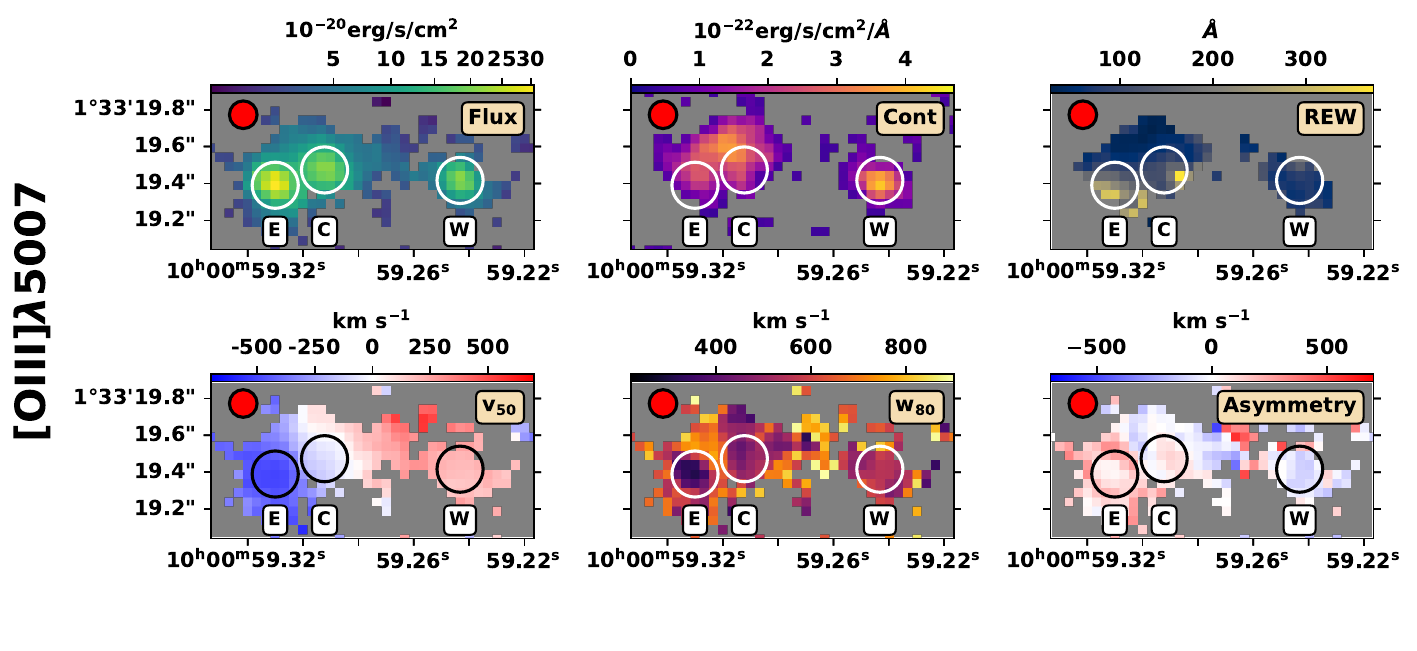}
\caption{Morpho-kinematic maps of \oiiib. We present the integrated line flux, continuum flux, rest-frame EW, $v_{50}$, $w_{80}$, and asymmetry per spaxel. The location of each component is marked with a hollow circle, and the fiducial PSF is represented by a red ellipse to the upper left.}
\label{morphomaps}
\end{figure*}

The fits are used to extract the integrated line flux ($F_L$), continuum flux ($S_C$), the rest-frame equivalent width, and relevant kinematic measures: $v_{50}$, $w_{80}$, and asymmetry$=|v_{50}-v_{90}|-|v_{50}-v_{10}|$ (e.g., \citealt{herv23}). We exclude spaxels with unrealistic kinematic measures caused by low signal (i.e., $|v_{50}|>700$\,km\,s$^{-1}$). The line and continuum fluxes are used to derive a rest-frame equivalent width. All acceptable fits are then used to create the morpho-kinematic maps shown in Figure \ref{morphomaps}.

The results of this analysis (Figure \ref{morphomaps}) are very similar to those of our two-Gaussian approach (Figure \ref{r2700mom_real}). The \oiiib emission exhibits peaks in three locations, the continuum emission peaks in \hzc and to the north of \hza and \hzb, a smooth east-west velocity gradient is observed, and the asymmetry pattern is maintained. Since the results of this more detailed kinematic analysis agrees with the two-Gaussian model, we use the two-Gaussian model in this work.

\bsp	
\label{lastpage}
\end{document}